\documentclass[twocolumn,tighten,times,twocolappendix,numberedappendix]{aastex631}

\usepackage{amsmath}
\usepackage{float}
\usepackage{color}
\usepackage{url}
\usepackage{etoolbox}
\usepackage{graphicx}

\usepackage[flushleft]{threeparttable}

\usepackage{placeins}

\hypersetup{breaklinks=true,colorlinks=true,linkcolor=blue,citecolor=blue,filecolor=blue,urlcolor=blue}
\makeatletter
\patchcmd{\NAT@citex}
  {\@citea\NAT@hyper@{%
     \NAT@nmfmt{\NAT@nm}%
     \hyper@natlinkbreak{\NAT@aysep\NAT@spacechar}{\@citeb\@extra@b@citeb}%
     \NAT@date}}
  {\@citea\NAT@nmfmt{\NAT@nm}%
   \NAT@aysep\NAT@spacechar\NAT@hyper@{\NAT@date}}{}{}

\patchcmd{\NAT@citex}
  {\@citea\NAT@hyper@{%
     \NAT@nmfmt{\NAT@nm}%
     \hyper@natlinkbreak{\NAT@spacechar\NAT@@open\if*#1*\else#1\NAT@spacechar\fi}%
       {\@citeb\@extra@b@citeb}%
     \NAT@date}}
  {\@citea\NAT@nmfmt{\NAT@nm}%
   \NAT@spacechar\NAT@@open\if*#1*\else#1\NAT@spacechar\fi\NAT@hyper@{\NAT@date}}
  {}{}

\usepackage{array}
\newcolumntype{C}[1]{>{\centering\let\newline\\\arraybackslash\hspace{0pt}}m{#1}}

\usepackage{xspace}

\newcommand{\orcidicon}{\includegraphics[width=0.26cm]{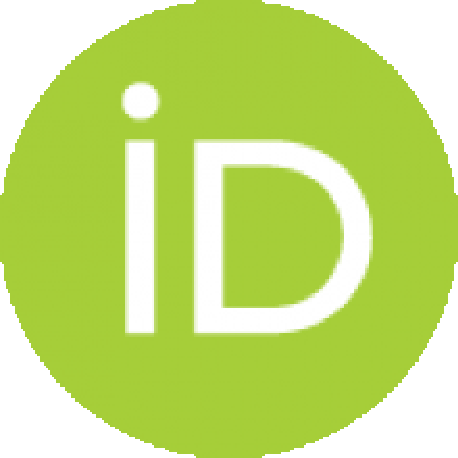}}

\newcommand{\orcidauthor}[1]{\href{https://orcid.org/#1}{\orcidicon}}

\def\araa{ARA\&A}
\def\apj{ApJ}
\def\apjl{ApJL}
\def\apjs{ApJS}
\def\ao{Appl.~Opt.}
\def\apss{Ap\&SS}
\def\aap{A\&A}

\def\mnras{MNRAS}

\def\pasp{PASP}

\def\ssr{Space~Sci.~Rev.}

\def\nat{Nature}

\def\physrep{Phys.~Rep.}

\newcommand{\xstar}{\textsc{xstar}\xspace}
\newcommand\mpixstar{\textsc{mpi\_xstar}\xspace}

\newcommand{\ionic}[2]{#1$\,${\scshape{#2}}\xspace}

\def\arcdeg{\hbox{$^\circ$}}

\makeatletter
\patchcmd{\frontmatter@RRAP@format}{(}{}{}{}
\patchcmd{\frontmatter@RRAP@format}{)}{}{}{}
\renewcommand\Dated@name{}
\makeatother

\renewcommand{\vec}[1]{\mathbf{#1}}

\shorttitle{Disk Reflection in NGC\,3783}
\shortauthors{Danehkar \& Brandt}
\graphicspath{{./}{figures/}}

\begin{document}

\title{Accretion-disk reflection in the Seyfert 1 galaxy NGC\,3783\\ as viewed by \textit{Chandra} grating spectroscopy}

\correspondingauthor{A.~Danehkar}

\author[0000-0003-4552-5997]{A.~Danehkar}
\affiliation{Eureka Scientific, 2452 Delmer Street, Suite 100, Oakland, CA 94602-3017, USA; \href{mailto:danehkar@eurekasci.com}{danehkar@eurekasci.com}}

\author[0000-0002-0167-2453]{W.~N.~Brandt}
\affiliation{Department of Astronomy and Astrophysics, The Pennsylvania State University, 525 Davey Lab, University Park, PA 16802, USA}
\affiliation{Institute for Gravitation and the Cosmos, The Pennsylvania State University, University Park, PA 16802, USA}
\affiliation{Department of Physics, The Pennsylvania State University, 104 Davey Laboratory, University Park, PA 16802, USA}



\date[]{\textit{\footnotesize Received 2024 July 8; revised 2025 January 8; accepted 2025 January 19}}

\begin{abstract}
In active galactic nuclei, X-ray illumination of the accretion disk around a supermassive black hole (SMBH) results in the production of the K$\alpha$ fluorescent line of iron, which provides insights into accretion physics and SMBH spins. In this work, we studied X-ray reflection from the accretion disk in the Seyfert 1 galaxy NGC\,3783 using all the data collected by the \textit{Chandra} High Energy Transmission Grating Spectrometer. We used hardness-ratio diagrams to distinguish between different spectral states and conducted spectral analysis of all the multi-epoch datasets, as well as the source in the observed spectral states. Our hardness analysis indicates that the source gradually evolved into a harder state (2013--2016) compared to the previous epochs (2000--2001). Our spectral modeling implies that the relativistically broadened iron emission from the innermost accretion disk is associated with a near-maximal SMBH spin ($a=0.98^{+0.02}_{-0.12}$) in all the datasets, even though the hard state was present in 17\% of them, and a consistent spin is also found in different spectral states. In addition, the narrow, bright Fe K$\alpha$ line from distant regions has an excess velocity of $620^{+80}_{-70}$\,km\,s$^{-1}$ relative to the rest frame, implying that some distant layers of the disk could be twisted. Our results suggest that, despite long-term changes in the X-ray brightness of NGC\,3783, likely caused by eclipsing material, the relativistic reflection can be constrained thanks to the substantial counts provided by multi-epoch observations, while a warped disk structure may be present. 
\end{abstract}

\keywords{\href{https://astrothesaurus.org/uat/16}{Active galactic nuclei (16)};
\href{https://astrothesaurus.org/uat/1388}{Relativistic disks (1388)};
\href{https://astrothesaurus.org/uat/1822}{X-ray sources (1822)}
\vspace{4pt}
}


\section{Introduction}
\label{ngc3783:introduction}



Supermassive black holes (SMBHs) serve as primary drivers for the most powerful events occurring within the central regions of active galactic nuclei (AGN), which are visible throughout different energy ranges of the electromagnetic spectrum, notably X-rays \citep[e.g.,][]{Shlosman1990}. The X-ray power-law-like continuum seen in AGN is believed to originate from the corona \citep[e.g.,][]{Galeev1979,Begelman1983} or at the base of a jet above the inner accretion disk \citep[e.g.,][]{Blandford1979,Bridle1984}, where Comptonization occurs. 
A fraction of the X-rays generated in the corona irradiate the optically dense disk beneath, resulting in X-ray reflection and iron K$\alpha$ emission visible above the power-law continuum at 6.4 keV due to the substantial abundance of iron \citep{Krolik1987}, which is composed of fluorescent and recombination emissions \citep[e.g.,][]{George1991,Fabian2000}. Furthermore, the typical Doppler effect, along with light bending and gravitational redshifting near the event horizon, causes a relativistic broadening of the fluorescent Fe K$\alpha$ line originating from the innermost part of the accretion disk \citep{Reynolds2003,Miller2007}. In particular, the gravitational redshift is mainly responsible for the elongated low-energy tail of the skew-symmetric broadened Fe K$\alpha$ line profile. Due to the proximity of the relativistic line-emitting region to the black hole, frame-dragging effects caused by a spinning black hole can have a significant impact on the emission profiles. Thus, the relativistically broadened X-ray Fe K$\alpha$ line serves as a potent indicator of the spin of SMBHs in AGN \citep[see reviews by][]{Brenneman2013,Reynolds2014,Reynolds2019}. 

\begin{table*}
\begin{center}
\caption[]{Observation log of NGC\,3783 with \textit{Chandra} ACIS-S/HETGS performed in the timed-exposure, faint mode.
\label{ngc3783:obs:log}}
\footnotesize
\begin{tabular}{lllllrrr}
\hline\hline\noalign{\smallskip}
Obs. ID  & Seq. No. & PI    & \multicolumn{1}{c}{Obs. Start (UTC)}     & \multicolumn{1}{c}{Obs. End (UTC)}        & Exp. (ks) & Count\,$^{\rm \bf a}$      & Cnt.\,Rate\,$^{\rm \bf a}$ \\
\noalign{\smallskip}
\hline
\noalign{\smallskip}
373  & 700045 & G.\,P. Garmire  & 2000 Jan 20, 23:33   & 2000 Jan 21, 16:20    &     56.43 &       12532 &      0.222 \\
2090 & 700280 & I.\,M. George   & 2001 Feb 24, 18:44   & 2001 Feb 26, 17:47    &    165.65 &       29910 &      0.181 \\
2091 & 700281 & I.\,M. George   & 2001 Feb 27, 09:17   & 2001 Mar 01, 09:09    &    168.85 &       30455 &      0.180 \\
2092 & 700282 & I.\,M. George   & 2001 Mar 10, 00:31   & 2001 Mar 11, 23:29    &    165.45 &       30849 &      0.186 \\
2093 & 700283 & I.\,M. George   & 2001 Mar 31, 03:36   & 2001 Apr 02, 02:48    &    166.13 &       44132 &      0.266 \\
2094  & 700284 & I.\,M. George  & 2001 Jun 26, 09:57   & 2001 Jun 28, 09:09    &    166.18 &       37285 &      0.224 \\
14991 & 702799 & W.\,N. Brandt  & 2013 Mar 25, 16:49   & 2013 Mar 26, 10:01    &     59.01 &        7466 &      0.127 \\
15626 & 702799 & W.\,N. Brandt  & 2013 Mar 27, 14:08   & 2013 Mar 28, 19:23    &    101.67 &       12104 &      0.119 \\
18192 & 703254 & L.\,W. Brenneman & 2016 Aug 22, 07:25   & 2016 Aug 23, 06:10    &     79.31 &       11090 &      0.140 \\
19694 & 703254 & L.\,W. Brenneman & 2016 Aug 25, 00:23   & 2016 Aug 25, 21:30    &     73.48 &        7946 &      0.108 \\
\noalign{\smallskip}\hline
\end{tabular}
\end{center}
\begin{tablenotes}
\footnotesize
\item[1]\textbf{Note.} $^{\rm \bf a}$ Source counts and rates of \textit{Chandra} ACIS-S/HETGS HEG data ($m = \pm 1$) over the energy range of 2--10\,keV.
\end{tablenotes}
\end{table*}


The nearby Seyfert 1 galaxy NGC\,3783  \citep[$z = 0.00896$;][]{Koss2022} hosts an AGN that is renowned for its relativistically broadened Fe K$\alpha$ line \citep{Brenneman2011,Reynolds2012}. The AGN possesses an exceptionally high level of optical brightness ($V \sim 13$\,mag, $L_{\rm bol}\approx 2.5 \times 10^{44}$\,erg\,s$^{-1}$) in the local universe at a distance of 38.5 Mpc \citep{Koss2022}. The SMBH mass of $2.5$--$2.8 \times 10^{7}$\,M$_{\odot}$ was determined from reverberation mapping studies \citep{Peterson2004,Bentz2021,GRAVITYCollaboration2021b}. However, there has been some debate regarding the SMBH spin rate. Some studies obtained a near-maximal spin with a lower limit of $0.89$ from relativistic Fe K$\alpha$ emission in \textit{Suzaku} observations \citep{Brenneman2011,Reynolds2012}. Nevertheless, other studies of the Fe K$\alpha$ line observed with \textit{Suzaku} proposed the possibility of a black hole rotating in the opposite direction or much slower, with an upper limit of $-0.04$ \citep{Patrick2011} and $0.24$ \citep{Patrick2012}, respectively. Moreover, two different methods used by \citet{Capellupo2017}, namely relativistic reflection and continuum-fitting, yield a wide range in the spin parameter, albeit with a high probability for a near-maximal spin if a disk wind is included in the reflection model.

Because the SMBH spin remains the same over human timescales, an appropriate combination of multi-epoch observations should have no impact on the spin measurement. To ensure that the reflection modeling is unaffected by X-ray variability, an AGN should be sufficiently bright and unblocked by obscuring materials to acquire the required signal-to-noise ratio with reliable photons to separate the relativistic reflection emission from the continuous emission of the corona or the base of the jet, as well as any outflow absorption features present in the host galaxy and its central region \citep[e.g.,][]{Tombesi2010,Tombesi2011,Laha2014,Laha2016,Danehkar2018}. Typically, it is necessary to have at least $2 \times 10^{5}$ net counts \citep{Guainazzi2006,Brenneman2013} or $1.5 \times 10^{5}$ net counts \citep{deLaCallePerez2010} within the energy range of 2--10 keV. However, with intricate soft absorption and obscuring outflows, the actual number of counts can be even higher. In particular, the Seyfert 1 galaxy NGC\,3783 is thought to have undergone some transient obscuration events caused by eclipsing outflow material near the X-ray source \citep{Mehdipour2017,Kriss2019}, in addition to showing persistent absorption features of an outflow \citep{Kaspi2002,Netzer2003}, so substantial counts would be beneficial for more accurate spin measurement. Combining observations from epochs under different spectral states provides enough photons to accurately constrain the long red wing of the fluorescent iron K$\alpha$ emission line, even though the X-ray brightness changes. In addition, proper spectral fitting can treat any effects on the red wing caused by the persistent outflowing absorption features.

In this work we explore all the \textit{Chandra} High Energy Transmission Grating Spectrometer (HETGS) archival data of NGC\,3783 to provide substantial photons for a robust constraint on the narrow Fe K$\alpha$ line and relativistically broadened Fe K$\alpha$ emission.  
Section~\ref{ngc3783:observation} describes \textit{Chandra} observations and data reduction. In Section~\ref{ngc3783:analyses}, we present our results from the hardness analysis, spectral modeling, and Bayesian analysis. A discussion is presented in Section~\ref{ngc3783:discussion}, followed by a conclusion in Section~\ref{ngc3783:conclusion}.

\vfill\break

\section{Observations and Data Reduction}
\label{ngc3783:observation}

NGC\,3783 was observed in total with an integration time of 333 hours and 56 minutes using the Advanced CCD Imaging Spectrometer S-array \citep[ACIS-S;][]{Garmire2003} and the High Energy Transmission Grating Spectrometer \citep[HETGS;][]{Canizares2005} aboard the \textit{Chandra} satellite \citep{Weisskopf2000,Weisskopf2002} from 2000 January 20 until 2016 August 25 during ten observing epochs.\footnote{The \textit{Chandra} datasets are available at doi:\href{https://doi.org/10.25574/cdc.200}{10.25574/cdc.200}.} There are two grating assemblies in the HETGS: the medium-energy grating (MEG) and the high-energy grating (HEG). The MEG covers 0.4--7 keV with a spectral resolution of 0.023\,{\AA}, while the HEG offers a resolution of 0.012\,{\AA} over 0.8--10 keV. The details of the observations are presented in Table~\ref{ngc3783:obs:log}, which in total have $2.2 \times 10^{5}$\,counts in the 2--10\,keV band with the HEG assembly ($m = \pm 1$ orders). As the effectiveness of the MEG starts to decline above 3 keV, we only used the HEG data for spectral analysis.

\begin{figure*}
\begin{center}
\includegraphics[width=0.92\textwidth, trim = 5 0 0 0, clip, angle=0]{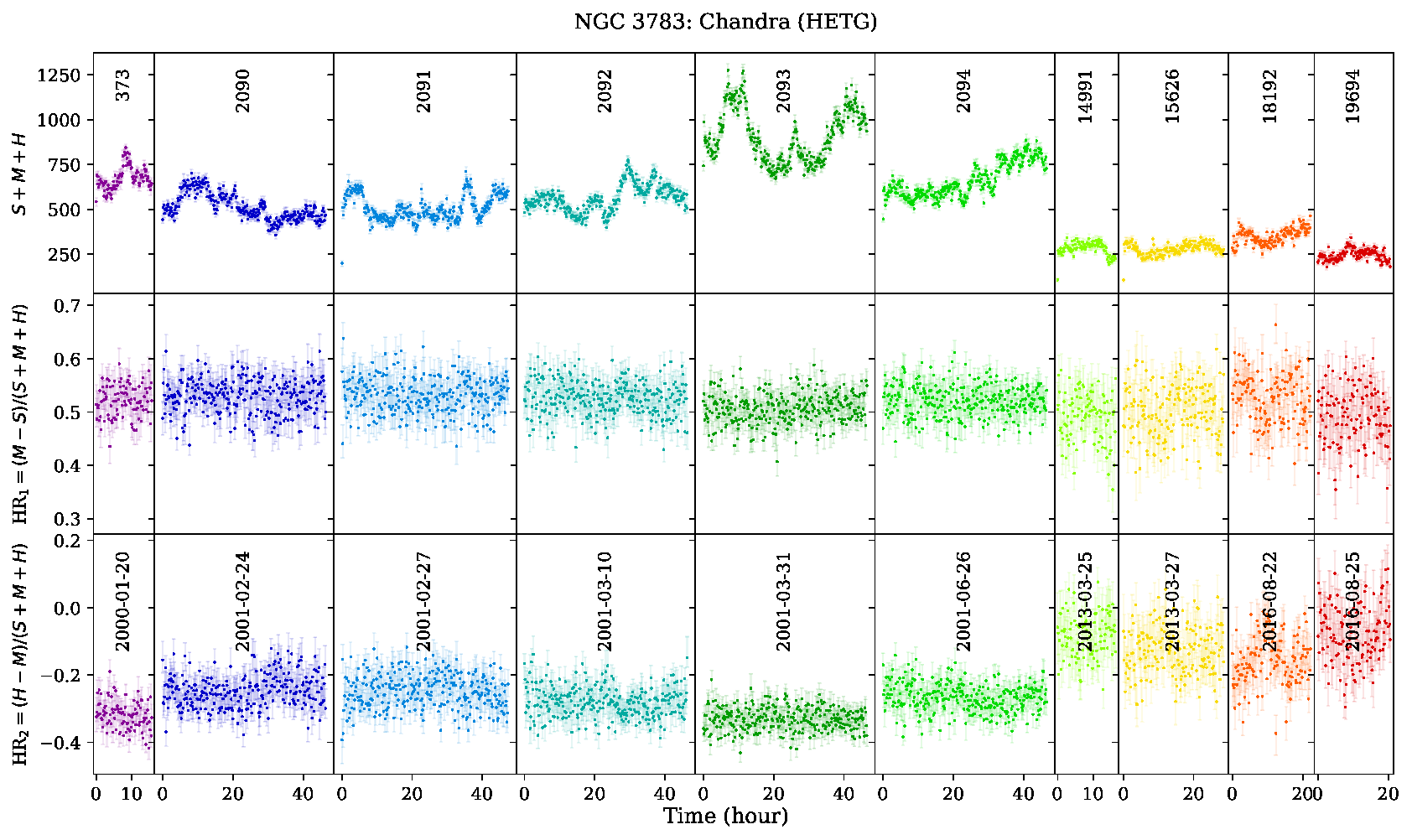}%
\end{center}
\caption{
The time series of NGC\,3783 in the broad bands ($S+M+H$; in counts), together with the corresponding hardness ratios ${\rm HR}_{1} = (M-S)/(S+M+H)$ and ${\rm HR}_{2} = (H-M)/(S+M+H)$ binned at 600 sec produced with \textit{Chandra} ACIS-S/HETGS observations.
\label{ngc3783:fig:lc:1}
}
\end{figure*}

We utilized \textsc{ciao} \citep[v\,4.15;][]{Fruscione2006} and its CALDB files (v4.10.2) to reduce the spectral data. The \textsc{ciao} task \textsf{chandra\_repro} was used to reprocess all events. This operation employs \textsf{acis\_process\_events} to generate the pulse height amplitude (PHA) data contained in the event files, \textsf{tgdetect2} to ascertain the zeroth-order centroid, and \textsf{tg\_resolve\_events} to resolve the spectral orders of each event. Moreover, it applies \textsf{tgextract} and \textsf{dmtype2split} to produce the dispersed MEG and HEG spectra in both the positive and negative first orders for each HETG event, in addition to the corresponding redistribution and response files with \textsf{mktgresp}. We also used the \textsc{ciao} task \textsf{dmextract} to produce time-binned light curves of the source and background from the first orders for different energy bands.

\begin{figure}
\begin{center}
\includegraphics[width=0.40\textwidth, trim = 0 0 0 0, clip, angle=0]{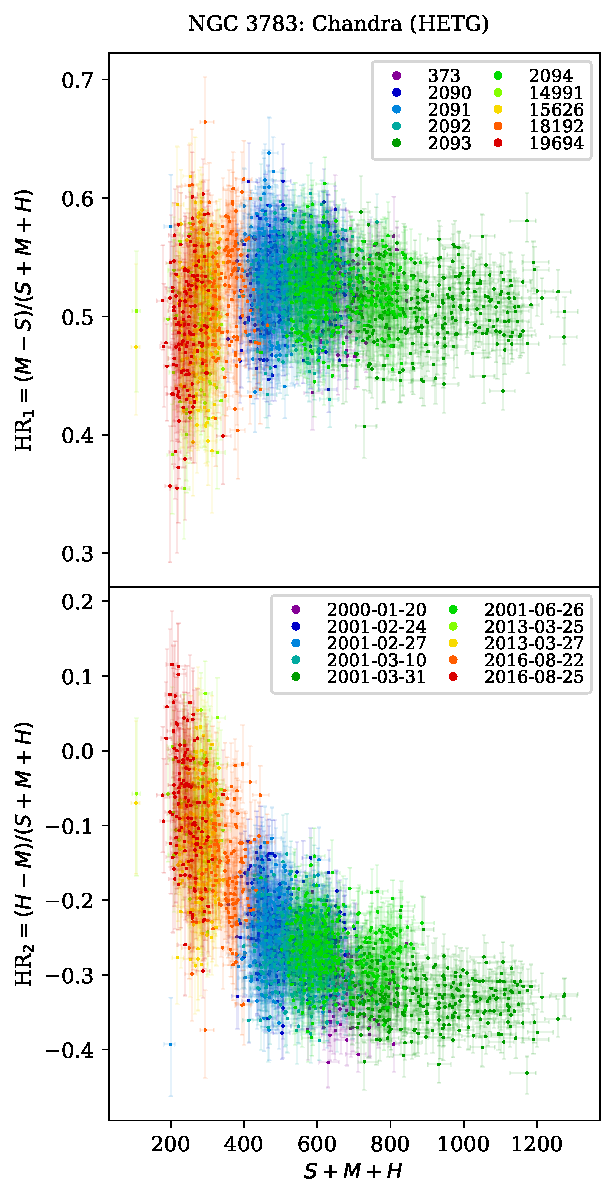}%
\end{center}
\caption{The hardness-ratio diagrams of NGC\,3783: the hardness ratios ${\rm HR}_{1} = (M-S)/(S+M+H)$ and ${\rm HR}_{2} = (H-M)/(S+M+H)$ plotted against the broad band ($S+M+H$; in counts) using the time series binned at 600 sec from \textit{Chandra} ACIS-S/HETGS observations, respectively.
\label{ngc3783:fig:hdr:1}
}
\end{figure}

\section{Analyses}
\label{ngc3783:analyses}

\subsection{Hardness Analysis}
\label{ngc3783:timing}

We generated light curves for the \textit{Chandra} HETGS observations of NGC\,3783, covering three specific energy ranges, namely the soft ($S$: 0.4--1.1 keV), medium ($M$: 1.1--2.6 keV), and hard ($H$: 2.6--8\,keV) bands, which allowed us to identify any X-ray transitions in NGC\,3783. In a way similar to \citet{Danehkar2024a}, we boosted the light curves of subsequent observations to compensate for the decrease in ACIS-S sensitivity over long timescales, using ratios found by integrating the effective area columns from the auxiliary response files (ARF) over the energy band of interest for the first observation relative to subsequent observations.

Hardness-ratio diagrams made using time series of various energy bands allow us to distinguish between different spectral states appearing throughout multi-epoch observations. To accomplish this, we calculated the hardness ratios using the light curves, as follows:
\begin{align}
\mathrm{HR}_{1} = \frac{M-S}{S+M+H}, ~~~~ & \mathrm{HR}_{2} = \frac{H-M}{S+M+H}. \label{eq_1}
\end{align}
This type of hardness analysis has been used to study the X-ray variability of various objects such as AGN, X-ray binaries, symbiotic stars, and other sources \citep[e.g.,][]{Finoguenov2002,Soria2003,Hong2004,Danehkar2024}. 
We can employ the hardness ratio $\mathrm{HR}_{2}$ to probe changes in the power-law continuum typically originating from the highly ionized corona or foot of a jet situated above the innermost accretion disk. The Fe K$\alpha$  emission line widened by Doppler and general-relativistic effects is present in the hard band, along with the non-relativistic bright, narrow iron K$\alpha$ line reflected from remote regions of the accretion disk (and beyond), although relativistic emission does not make a substantial contribution to the total counts. 
The hardness ratio ${\rm HR}_{1}$ exhibits variation in the medium band, which typically includes absorption lines and edges from warm absorbers \citep[e.g.,][]{Turner1993,Reynolds1997,George1998,Kaspi2002}. This ratio may also correspond to changes in the soft excess, which often includes multi-thermal black-body emissions from the accretion disk \citep[e.g.,][]{Abramowicz1988}, as well as mildly Compton-scattered emissions from a cooler component of the corona \citep[e.g.,][]{Hubeny2001,Wilkins2015}. 

\begin{table*}
\begin{center}
\caption[]{Hardness statistical analysis.
\label{ngc3783:stat:result}}
\begin{tabular}{lcccccccc}
\hline\hline\noalign{\smallskip}
Parameter     &$\eta$      &${\eta}_{\rm norm}$    &\multicolumn{2}{c}{Lilliefors}   &\multicolumn{2}{c}{Anderson--Darling}   &\multicolumn{2}{c}{Shapiro--Wilk}   \\
     &      &    &$D$      &$p$-value   &$A^2$     &$p$-value   &$W$     &$p$-value  \\
\noalign{\smallskip}
\hline
\noalign{\smallskip}
\multicolumn{9}{c}{All datasets} \\
\noalign{\smallskip}
$S$+$M$+$H$        & $    0.05 $ & ${    2.00 \pm 0.09 }$ & $    0.057 $ & $    0.001 $ & $   14.802 $ & $    0.000 $ & $    0.964 $ & $    0.000 $ \\
HR$_1$       & $    1.73 $ & ${    2.00 \pm 0.09 }$ & $    0.041 $ & $    0.001 $ & $    5.191 $ & $    0.000 $ & $    0.987 $ & $    0.000 $ \\
HR$_2$        & $    0.59 $ & ${    2.00 \pm 0.09 }$ & $    0.094 $ & $    0.001 $ & $   35.781 $ & $    0.000 $ & $    0.944 $ & $    0.000 $ \\
\noalign{\smallskip}
\multicolumn{9}{c}{High/soft (Group\,1)} \\
\noalign{\smallskip}
$S$+$M$+$H$        & $    0.07 $ & ${    2.00 \pm  0.10 }$ & $    0.121 $ & $    0.001 $ & $   40.165 $ & $    0.000 $ & $    0.906 $ & $    0.000 $ \\
HR$_1$        & $    1.87 $ & ${    2.00 \pm  0.10 }$ & $    0.015 $ & $    0.613 $ & $    0.397 $ & $    0.369 $ & $    0.999 $ & $    0.816 $ \\
HR$_2$        & $    1.28 $ & ${    2.00 \pm  0.10 }$ & $    0.020 $ & $    0.171 $ & $    1.147 $ & $    0.005 $ & $    0.996 $ & $    0.000 $ \\
\noalign{\smallskip}
\multicolumn{9}{c}{Low/hard (Group\,2)} \\
\noalign{\smallskip}
$S$+$M$+$H$        & $    0.43 $ & ${    2.00 \pm 0.17 }$ & $    0.044 $ & $    0.024 $ & $    2.028 $ & $    0.000 $ & $    0.985 $ & $    0.000 $ \\
HR$_1$        & $    1.66 $ & ${    2.00 \pm 0.17 }$ & $    0.033 $ & $    0.235 $ & $    0.597 $ & $    0.118 $ & $    0.996 $ & $    0.297 $ \\
HR$_2$        & $    1.54 $ & ${    2.00 \pm 0.17 }$ & $    0.025 $ & $    0.635 $ & $    0.278 $ & $    0.648 $ & $    0.998 $ & $    0.926 $ \\
\noalign{\smallskip}\hline
\end{tabular}
\end{center}
\begin{tablenotes}
\footnotesize
\item[1]\textbf{Notes.} There is autocorrelation in the time series if the von Neumann ratio ($\eta$) is outside the normal von Neumann ratio (${\eta}_{\rm norm}$). The randomness of the data is examined using $p$-values from the Lilliefors ($D$), Anderson--Darling ($A^2$), and Shapiro--Wilk ($W$) statistical methods.
\end{tablenotes}
\end{table*}

The light curves of NGC\,3783 are shown in Figure~\ref{ngc3783:fig:lc:1}, corresponding to the broad band ($S+M+H$) binned at 600\,sec. They were corrected for background by applying the Bayesian Estimator for Hardness Ratios \citep[BEHR;][]{Park2006} to the source and background time series extracted from the HETGS observations. The plot also includes the corresponding hardness ratios, ${\rm HR}_{1}$ and ${\rm HR}_{2}$. The uncertainties in the light curves and hardness ratios are based on the BEHR Monte Carlo simulations. Although all the time series exhibit stochastic low-amplitude fluctuations, we notice that the source gradually became brighter from 2000 February 24 until 2001 March 31, followed by a decrease in the X-ray brightness on 2001 June 26. It can be seen that the source was much fainter in 2013 and 2016. It also slowly experienced a moderate increase in brightness from 2013 March 25 until 2016 August 22, but slightly fainter on 2016 August 25. The hardness ratio ${\rm HR}_{1}$ remained approximately of the same magnitude in all observations, albeit with higher variability in 2013 and 2016. Similarly, the hardness ratio ${\rm HR}_{2}$ experienced increased fluctuations in 2013 and 2016, but with a higher average level. This implies that the source hardness was higher in 2013 and 2016, whereas there was higher variability in the medium band, which could be associated with absorbers present in that band.

Figure~\ref{ngc3783:fig:hdr:1} presents the hardness-ratio diagrams constructed by graphing the aforementioned hardness ratios against the broad band ($S+M+H$), which aid in the detection of possible X-ray flares or obscurers. It appears that the X-ray source moved to \textit{low/hard} and \textit{high/soft} spectral states on a year-scale, in addition to hourly small-amplitude stochastic variations. According to the $\mathrm{HR}_{2}$ diagram, the source was in a faint/hard state in 2013 and 2016, whereas it was in a bright/soft state in 2000 and 2001. NGC\,3783 is believed to experience so-called transient obscuration events produced by an eclipsing 
outflow near the X-ray source \citep{Mehdipour2017,Kriss2019}. This type of transient obscuration has been identified in other AGNs such as NGC\,5548 \citep{Kaastra2014}, NGC\,985 \citep{Ebrero2016}, NGC\,3227 \citep{Turner2018}, Mrk\,335 \citep{Longinotti2019,Parker2019}, and ESO\,33-2 \citep{Walton2021}.
Alternatively, it might be associated with increases in X-ray flaring coronae in the innermost central regions. PDS 456, which contains nearly relativistic outflows \citep{Reeves2018,Boissay-Malaquin2019}, is one of the AGNs with long- and short-term X-ray variability caused by coronal X-ray flares \citep{Matzeu2017a,Reeves2021}. We should also mention NGC\,3516, which is thought to contain partially variable X-ray obscuration \citep{Oknyansky2021}, although \citet{Mehdipour2022} suggested that changes in the ionizing source induce increased X-ray absorption caused by eclipsing outflows, mimicking the presence of obscuration events.

To assess the extent of variability, we conducted statistical analyses on the time series. We not only performed our analyses on all the datasets, but also on the datasets divided into two groups: Group 1, which represents the high/soft state (2000--2001), and Group 2, which is regarded as the low/hard state (2013--2016). To identify autocorrelation, we computed the von Neumann ratio \citep{vonNeumann1941}, which is the mean squared successive difference relative to the variance, as follows:
\begin{equation}
\eta=\frac{\delta^2}{\sigma^2}=\frac{\sum_{i=1}^{n-1}(x_{i+1}-x_i)^2/(n-1)}{\sum_{i=1}^{n}(x_i-\bar{x})^2/(n-1)},\label{eq_4}
\end{equation}
where $i$ is the index of each point, $n$ is the total number of points, and $\bar{x}=\sum_{i=1}^{n}x_i/n$ is the mean. The von Neumann ratio of a normal probability distribution is expected to be ${\eta}_{\rm norm}=2n/(n-1) \sim 2$  \citep{Young1941}. To evaluate autocorrelation, we obtained the confidence limit of the normal von Neumann ratio at the level of $\alpha=0.05$. If the von Neumann ratio is within the confidence limit of the normal von Neumann ratio, there is no autocorrelation. However, if the ratio is outside these confidence limits and closer to 0 or 4, it indicates a positive or negative autocorrelation, respectively.

To examine the hypothesis of normality (randomness) in the data, we utilized three statistical methods to determine whether the data follows a normal (random) distribution: the Lilliefors statistic, the Anderson--Darling method, and the Shapiro--Wilk method. The Lilliefors technique \citep{Lilliefors1967}, which is a modified version of the Kolmogorov-Smirnov method, involves using the estimated mean and variance from the data to assess normality. This statistic provides the largest difference ($D$) between the empirical distribution function (EDF) of a sample and the cumulative distribution function (CDF) of a normal distribution. The Anderson--Darling test \citep{Anderson1952} finds the squared difference ($A^2$) between the EDF and CDF and focuses on the tails of the distribution. The Shapiro--Wilk statistic \citep[$W$;][]{Shapiro1965} is a method that relies on order statistics, predicted values of order statistics of independent and normal variables, and the covariance of these order statistics. We calculated the Lilliefors and Anderson-Darling statistics using the respective procedures from the \textsf{Statsmodels} package \citep{Seabold2010}, and used the appropriate function from the \textsf{SciPy} package \citep{Virtanen2020} for the Shapiro-Wilk statistic. If the $p$-value of these statistics is less than or equal to the significance level of $\alpha=0.05$, we reject the hypothesis of normality, meaning that there is non-random variability. The Shapiro--Wilk test is the most statistically powerful method, whereas the Anderson--Darling test is more effective when dealing with a distribution having a distinct peak and tails that end abruptly \citep{Stephens1974,Razali2011}.

Table~\ref{ngc3783:stat:result} presents the results of the variations in the broad band ($S+M+H$) and the hardness ratio for NGC\,3783, derived from various methodologies. It can be seen that the von Neumann ratios generally suggest autocorrelation in time series since they are outside the normal von Neumann domains (${\eta}_{\rm norm}$). However, positive autocorrelation is stronger (closer to 0) in the broad band ($S+M+H$) and the ${\rm HR}_{2}$ hardness ratio in all datasets, as well as $S+M+H$ under different spectral states. In addition, the Anderson--Darling and Shapiro--Wilk methods can determine whether random distributions exist in the time series. The $p$-values from these statistical methods are below the significance threshold of 0.05, leading to rejection of the normality hypothesis. In the case of all the observations, there are no normal distributions, so the spectral transitions in the broad band and hardness ratios, as seen in the hardness-ratio diagrams (Fig.\,\ref{ngc3783:fig:hdr:1}), are considered to be statistically significant. When considering the time series of the source during the high/soft state (Group 1), there are statistically significant fluctuations in the broad band ($S+M+H$) and the ${\rm HR}_{2}$ hardness ratio, whereas the ${\rm HR}_{1}$ hardness ratio has a normal (random) distribution. However, the Lilliefors test suggests a normal distribution in ${\rm HR}_{2}$ in Group 1. In the case of the source in the low/hard state (Group 2), the variations in the broad band appear to be non-random, while the hardness ratios demonstrate random fluctuations. 

In summary, the broad-band light curves change in ways that are statistically significant across all the observations and under different spectral states. There are also spectral transitions in the hardness ratios across all the observations and the ${\rm HR}_{2}$ of the observations in the high/soft state based on the Anderson--Darling and Shapiro--Wilk tests. Our statistical analyses of the X-ray variability indicate that there is a statistically significant hardness transition over the course of all observations, so it is essential to consider the implications of different spectral states (high/soft and low/hard) for spin measurement.



\subsection{Spectral Modeling}
\label{ngc3783:spec}

We conducted time-averaged spectral analysis on all HEG $m = \pm 1$ orders of the available HETGS data. 
We also categorized the observations into two groups under different spectral states: high/soft (Group\,1; 2000--2001) and low/hard (Group\,2; 2013--2016). These three datasets (all of them and two groups) were considered for our data analysis.

We used the Interactive Spectral Interpretation System\footnote{\href{https://space.mit.edu/asc/isis/}{https://space.mit.edu/asc/isis/}} \citep[\textsc{isis} v.\,1.6.2-51;][]{Houck2000}, which provides access to various spectral models included in the X-ray spectral fitting package \textsc{xspec} \citep[v.\,12.13.0;][]{Arnaud1996}.  We binned all the X-ray datasets such that there are at least 25 counts per energy bin. We then regridded them to match each other's energy bins and combined them based on the spectral state using the \textsc{isis} standard functions. 
In addition, the datasets were evaluated with the aid of a cross-normalization constant across different spectral states (\textsc{xspec} model \textsf{constant}).\footnote{This is done with the use of \textsf{constant(Isis\_Active\_Dataset)} in \textsc{isis}.}  This approach, which allows data-dependent fitting of a spectral model to different instruments and epochs, has been used in the spectral modeling of X-ray observations of AGNs \citep[e.g.,][]{Porquet2018,Masini2019,Sengupta2023}. Our choice to use a state-dependent constant is based on the fact that the X-ray variability in most Seyfert I AGNs predominantly ($\gtrsim95$\%) originates from changes in the normalization of the source continuum, as demonstrated by principal component analysis (e.g., \citeauthor{Parker2014} \citeyear{Parker2014}, \citeyear{Parker2015}; see review by \citeauthor{Danehkar2024b} \citeyear{Danehkar2024b}).
We restricted our spectral analysis to the energy range of 2--7.5\,keV in the rest frame, which contains the skew-symmetric broadened Fe K$\alpha$ line profile and excludes the soft excess. The fitting procedure was implemented using the chi-square ($\chi^2$) statistical method, in conjunction with the Levenberg-Marquardt (\textsf{mpfit}) optimization algorithm \citep{More1978,Garbow1980}.\footnote{The \textsc{isis} implementation of \textsf{mpfit} is based on the function translated from Fortran to C language by S.~L.~Moshier and improved by \citet{Markwardt2009}.} 
In our spectral modeling, we assumed the elemental abundances of the ISM from \citet{Wilms2000} as solar values.

Before modeling the reflection components, we explored a spectral model that did not include relativistic reflection emission. This model had a powerlaw component and a distant reflection component made with \textsf{xillverCp} \citep{Garcia2010,Garcia2011,Garcia2013}, in addition to the Galactic absorbing column and a warm absorber explained later in this section. We found that this model produces a nonphysical photon index of approximately $1.4$--$1.5$, which falls short of the typical range of $\Gamma \sim 1.6$--$2.2$ observed in Seyfert 1 AGN and disagrees with $\Gamma$\,$=$\,$1.6$--$1.8$ reported by \citet{Brenneman2011} and \citet{Reynolds2012}, indicating a physically incorrect reproduction of the curvature in the continuum.
Therefore, to describe the continuum, we proceeded with our spectral model, in which a \textsf{relxillCp} component \citep[v2.3.1;][]{Garcia2014,Dauser2014,Dauser2016} produces relativistic reflection emission along with a primary X-ray source internally created with the thermally Comptonized continuum model \citep[\textsf{nthcomp};][]{Zdziarski1996,Zycki1999} for self-consistent photoionization calculations of relativistic reflection. The non-relativistic reflection \textsf{xillverCp} component comprises only reflection (by setting a negative unit in the reflection fraction). However, its source photon index ($\Gamma$) was linked to that of \textsf{relxillCp}. We set the $z$ variable of \textsf{xillverCp} as a free parameter to improve statistical fit. The redshift parameter of \textsf{relxillCp} was frozen at the rest-frame redshift of the host galaxy ($z=0.009755$). 
The innermost and distant regions of the accretion disk are expected to be illuminated by the same X-ray source defined by the photon index. Moreover, a uniform equatorial distribution of matter in the accretion disk results in the same inclination angle across different regions. Accordingly, the photon index and inclination of the \textsf{xillverCp} component were tied to those of \textsf{relxillCp}. Assuming a chemically homogeneous distribution of elements in the accretion disk, the iron abundance parameter of \textsf{xillverCp} was also linked to that of the \textsf{relxillCp} component. 

\setcounter{figure}{2}
\begin{figure*}
\begin{center}
\includegraphics[width=0.58\textwidth, trim = 0 0 0 0, angle=0]{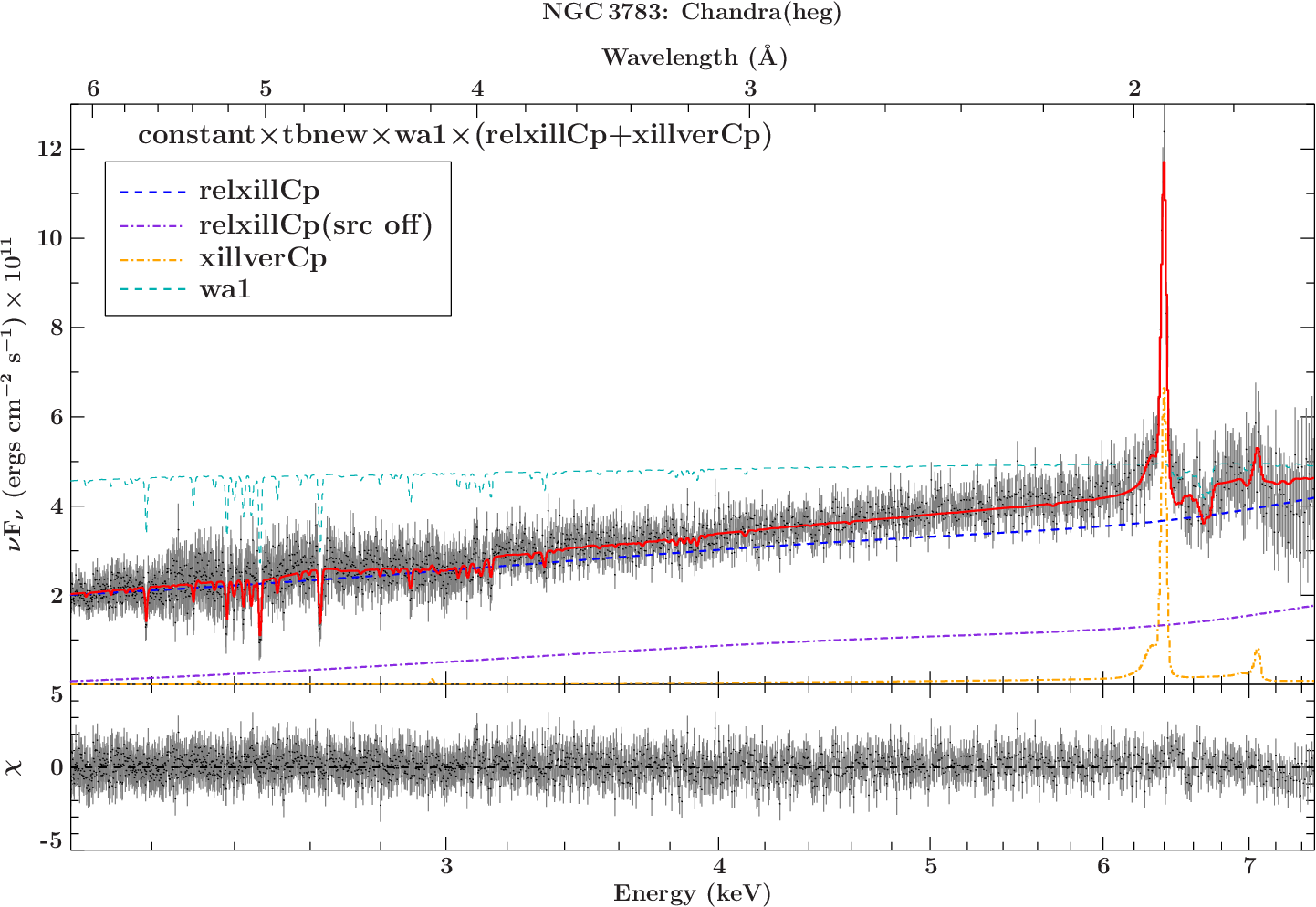}\\\smallskip
\includegraphics[width=0.58\textwidth, trim = 0 0 0 0, angle=0]{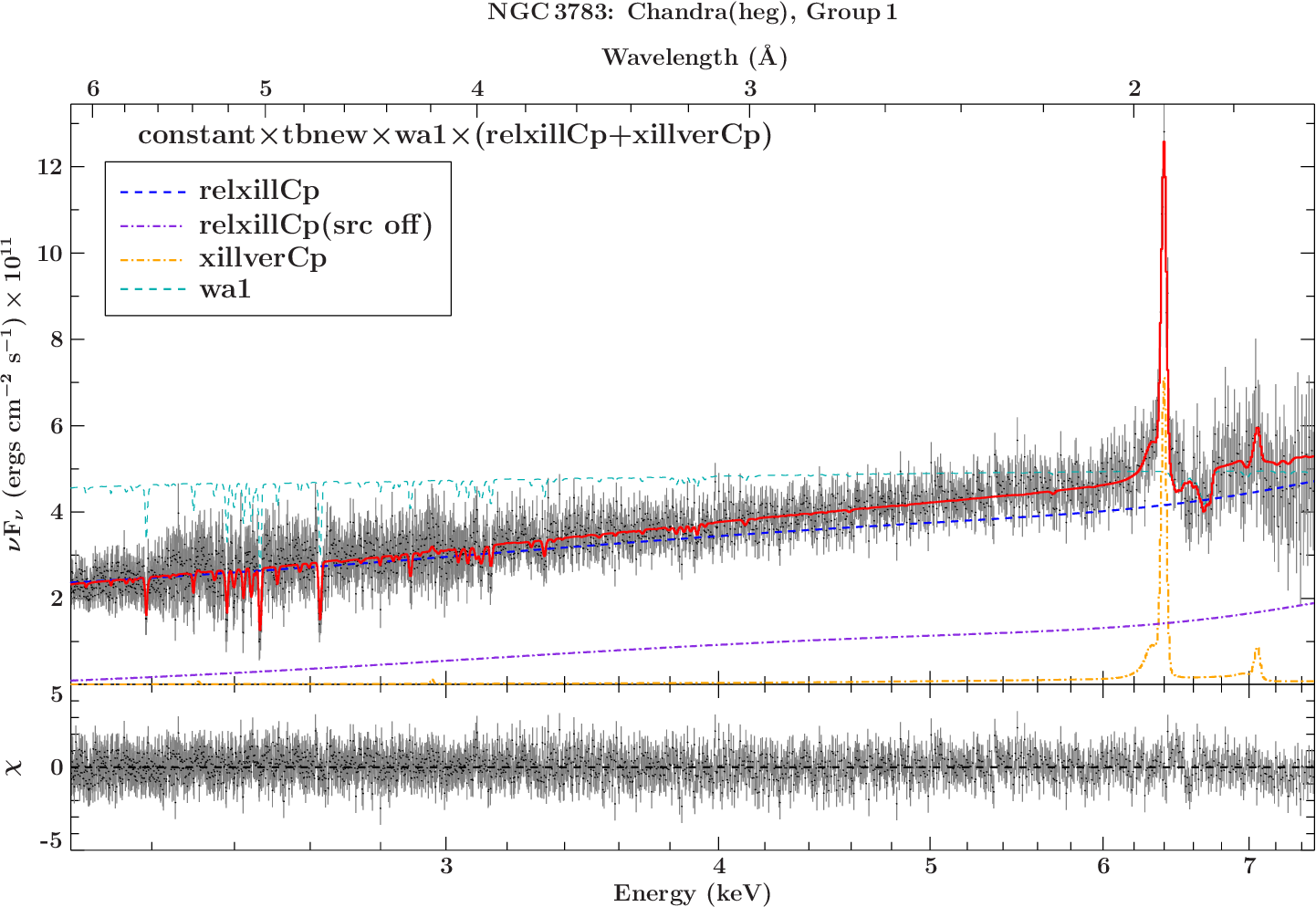}\\\smallskip
\includegraphics[width=0.58\textwidth, trim = 0 0 0 0, angle=0]{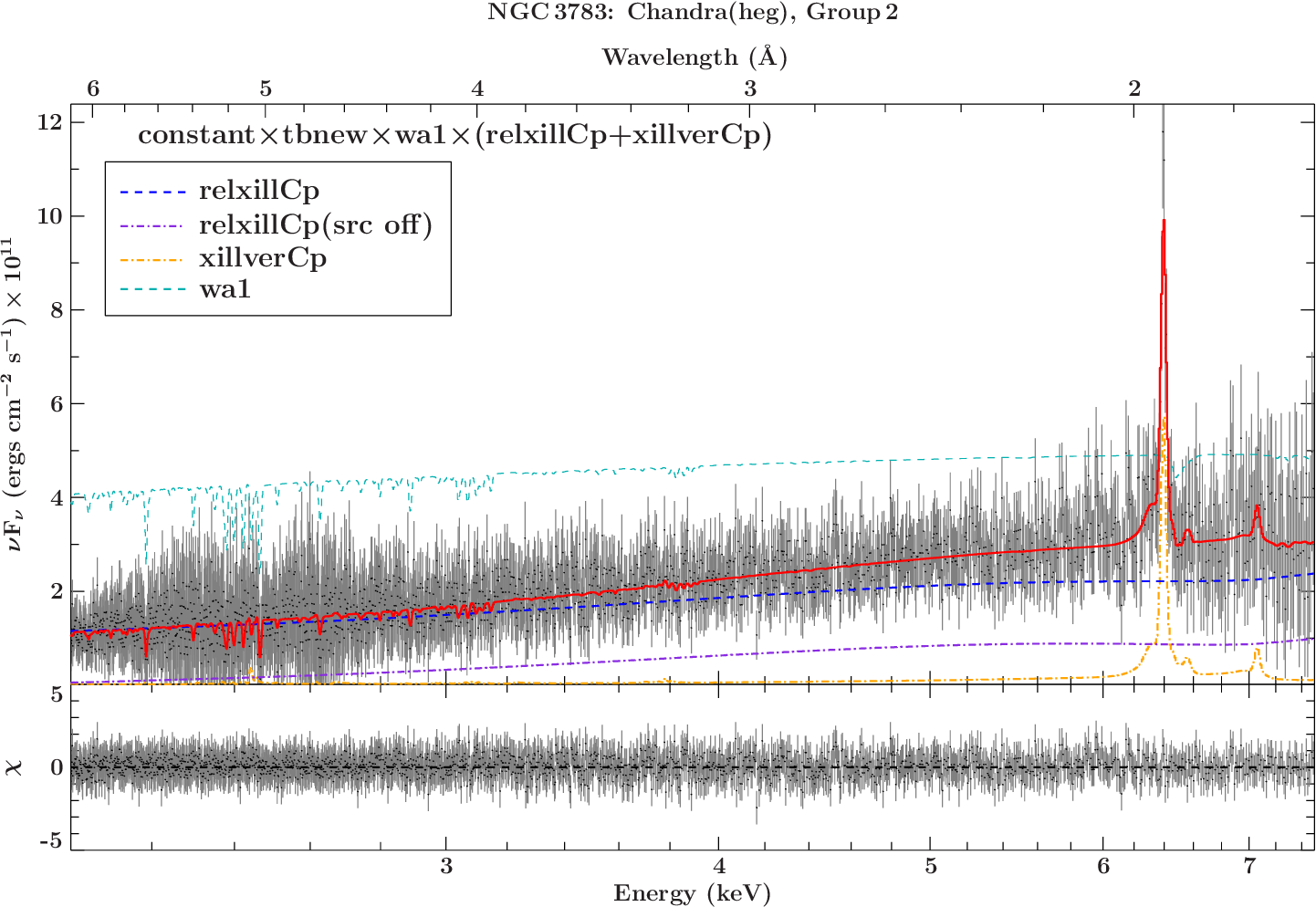}
\end{center}
\caption{The HEG data ($m=\pm1$ orders) of all the \textit{Chandra} HETGS observations (top panel) of NGC\,3783 fitted with the spectral model (red solid line) \textsf{constant}\,$\times$\,\textsf{tbnew}\,$\times$\,\textsf{wa1}\,$\times$\,(\textsf{relxillCp}\,$+$\,\textsf{xillverCp}), as well as Group 1 (high/soft; middle) and Group 2 (low/hard; bottom) observations. The contribution of each component to the model is separately plotted: \textsf{relxillCp} with the direct continuum (blue dashed lines), \textsf{relxillCp} (src off; purple dashed lines) without the direct continuum, \textsf{xillverCp} (orange dash-dotted line), \textsf{zpowerlw} (purple color dotted lines), and \textsf{wa1} (cyan dashed lines;  arbitrary scale). The lower part of each panel displays the standardized $\chi$ residuals.
\label{ngc3783:fig:model}
}
\end{figure*}

We included two components to describe the Galactic foreground absorption and X-ray absorption features of the host galaxy. To account for foreground absorption that mostly affects the soft excess, we incorporated a \textsf{tbnew} component \citep{Wilms2000}, whose column density was fixed at the weighted Galactic column density of \ionic{H}{i} ($N_{\rm H\,I}= 9.91\times 10^{20}$\,cm$^{-2}$) from the UK Swift Science Data Centre\footnote{\href{https://www.swift.ac.uk/analysis/nhtot/}{https://www.swift.ac.uk/analysis/nhtot/}} \citep[UKSSDC;][]{Willingale2013}. Our analysis indicated the presence of a soft X-ray absorption component, which is well-known and has been extensively studied in previous works \citep[e.g.,][]{Kaspi2002,Netzer2003}. Without a warm absorber component, we obtained $\chi^2/{\rm d.o.f}=1.39$ (and 1.40 excluding also \textsf{tbnew}) for the model fitted to all the first-order HEG data, whereas its inclusion resulted in $\chi^2/{\rm d.o.f}=0.84$. The absorbing outflow component is modeled using the \xstar code \citep[v\,2.59;][]{Kallman1996,Kallman2001,Kallman2004};
for a warm absorber (\textsf{wa1}) defined by the column density ($N_{\rm H}$), ionization parameter ($\log \xi$), and outflow velocity ($v_{\rm out}$) with respect to the rest frame. We produced a grid of $9 \times 11$ \xstar models in the $N_{\rm H}$--$\xi$ space, sampling the column density with 9 logarithmic intervals (from $\log N_{\rm H}=20$ to $24$\,cm$^{-2}$ with a logarithmic interval step of $0.5$) and the ionization parameter with 11 intervals (from $\log\xi=0$ to $5$ erg\,cm\,s$^{-1}$ with an interval step of $0.5$), assuming a gas density of $n=10^{12}$\,cm$^{-3}$, a turbulent velocity of $v_{\rm turb}=200$\,km\,s$^{-1}$, elemental abundances fixed to solar values, and a power-law ionizing source with a spectral index of $\Gamma = 1.6$ ($\alpha=-\Gamma+1$ in \xstar). 
To speed up grid construction, we utilized \mpixstar \citep{Danehkar2018a}, which allows multiple  \xstar programs to run in parallel with the aid of the OpenMPI library \citep{Gabriel2004a}. The multiplicative tabulated model of absorption spectra imprinted onto continua (\textsf{xout\_mtable.fits}) was used for spectral analysis of the X-ray absorption lines and edges.

The count data are fitted to each spectral model by adjusting the normalization parameters and folding the model through the instrument response function using the standard \textsc{isis} functions, which provides prior knowledge for further Bayesian statistics. All HEG data ($m=\pm1$ orders) of the ACIS-S/HETGS observations fitted with the model are shown in Figure~\ref{ngc3783:fig:model}, along with the standardized $\chi$ residuals, $\chi = ({\rm data} - {\rm model})/\Delta$, where $\Delta$ is the statistical error. The model components are also shown with different colors and line styles, which aid in distinguishing the contributions from the innermost and distant regions of the accretion disk (\textsf{wa1} component on an arbitrary scale). 

Table~\ref{ngc3783:model:params} presents the values derived in our spectral modeling, as well as those constrained by Bayesian statistics denoted by ``(m)'' described in the next section.  
The confidence limits at 90\% for the parameters were determined using the \textsc{isis} standard procedure for confidence limits (\textsf{conf\_loop}). The SMBH spin ($a$), inclination angle ($i$), inner radius ($R_{\rm in}$, set to the innermost stable circular orbit (ISCO) $R_{\rm ISCO}$), and emissivity profile of the accretion disk all determine the relativistic emission \citep[e.g.,][]{Reynolds2003}. Following the approach widely used in the literature \citep[e.g.,][]{Ding2022,Vaia2024}, we fixed the outer radius  of the accretion disk ($R_{\rm out}$) to $400 R_g$ since it is insensitive to spectral fitting, where $R_g \equiv GM/c^2$ is the gravitational radius. The ISCO is a function of the spin, e.g., $R_{\rm ISCO}|_{a=1}=R_g$ and  $R_{\rm ISCO}|_{a=0}=6R_g$, so the inner radius is altered by the spin. These inner and outer radii are identical to those assumed in previous X-ray spin studies of NGC\,3783 \citep{Brenneman2011,Reynolds2012}. 

The emissivity profile in \textsf{relxillCp} is described by a broken powerlaw with the form $\propto r^{-q_{\rm in}}$ between $R_{\rm in}$ and $R_{\rm br}$ and $\propto r^{-q_{\rm out}}$ between $R_{\rm br}$ and $R_{\rm out}$. Here, $q_{\rm in}$ and $q_{\rm out}$ are the inner and outer emissivity indices of the disk, respectively, and $R_{\rm br}$ is the break radius. To simplify the emissivity profile, we adopted an approach similar to that used by \citet{Victoria-Ceballos2023}, in which $R_{\rm br}$ was set to $(R_{\rm out}-R_{\rm in}|_{a=1})/2+R_{\rm in}|_{a=1}$ and $q_{\rm out}$ was tied to $q_{\rm in}$. We assumed a constant density throughout the disk, with a value of $10^{15}$\,cm$^{-3}$, which is commonly assumed for accretion disks in AGNs \citep[e.g.,][]{Jiang2019,Du2024}. We did not fix the reflection fraction of \textsf{relxillCp} ($f_{\rm refl,rel}$), which corresponds to the intensity of the coronal radiation illuminating the disk with respect to the coronal radiation directly reaching the observer \citep{Dauser2014}. As seen in Table~\ref{ngc3783:model:params}, the reflection fraction of the relativistic component slightly increased when the source was in the low/hard spectral state, whereas the normalization factor ($K_{\rm rel}$) decreased. In addition, the radial emissivity profile ($r^{-q}$) of the accretion disk was characterized by a sharper drop (higher $q$) in ionization, as well as a lower ionization parameter ($\log \xi_{\rm xil}$) in the distant region (\textsf{xillverCp}), during the high/soft spectral state. Furthermore, the low level of ionization in the distant region explains the slightly lower normalization factor ($K_{\rm xil}$) of \textsf{xillverCp} when the source was in the high/soft state. 

To improve the statistics, we also relaxed the constraint on the redshift of the distant reflection line (\textsf{xillverCp}), resulting in an excess velocity of $v_{\rm shift} = 620^{+80}_{-70}$\,km\,s$^{-1}$ relative to the rest frame of the host galaxy. This indicates that some parts of the non-relativistic accretion disk in the central regions might be warped in the line of sight, similar to NGC\,4151 \citep{Miller2018,XRISMCollaboration2024} and NGC\,5128 \citep[Centaurus\,A;][]{Bogensberger2024}.
Considering the confidence constraints, the excess velocity ($v_{\rm shift}$) of the non-relativistic Fe K$\alpha$ line is approximately similar in all groups, although its mean value was higher in the low/hard state. The outflow velocity range ($v_{\rm out}$) of the warm absorber (\textsf{wa1}) was roughly the same in the different dataset groups.

\subsection{Bayesian Statistics}
\label{ngc3783:mcmc}

We used the results of our spectral analysis to establish posterior constraints on the best-fitting parameters using Bayesian statistical analysis. We employed an MPI parallelization implementation\footnote{\href{https://github.com/mcfit/slmpi_emcee}{https://github.com/mcfit/slmpi\_emcee}} of the \textsc{isis} \textsf{emcee} hammer routine \citep{Nowak2016}, which is an S-Lang implementation of the Ensemble samplers prescribed by \citet{Goodman2010}. 
We performed 1,000,000 sampling iterations using $2000$ walker steps, $5$ free parameters, and $100$ random walkers (i.e., 500 iterations per walker step), assuming Gaussian statistics. However, half of these iterations were sufficient to achieve convergence in Group 2 (low/hard). Markov chain Monte Carlo (MCMC) sampling is convergent if the acceptance rate is between 0.2 and 0.5 \citep{Gelman1996}. However, the ``curse of dimensionality'' \citep{Bellman1957}, also known as the Hughes phenomenon \citep{Hughes1968} and the peaking phenomenon \citep{Trunk1979}, can emerge from a high number of parameters in Bayesian statistics, resulting in higher 
probabilities of errors and continuously non-convergent Markov chains. To avoid the curse of dimensionality, we restricted our Bayesian analysis to the key parameters in MCMC sampling, which allowed us to achieve convergence in our MCMC simulations.

\setcounter{figure}{3}
\begin{figure}
\begin{center}
\includegraphics[width=0.48\textwidth, trim = 0 0 0 0, angle=0]{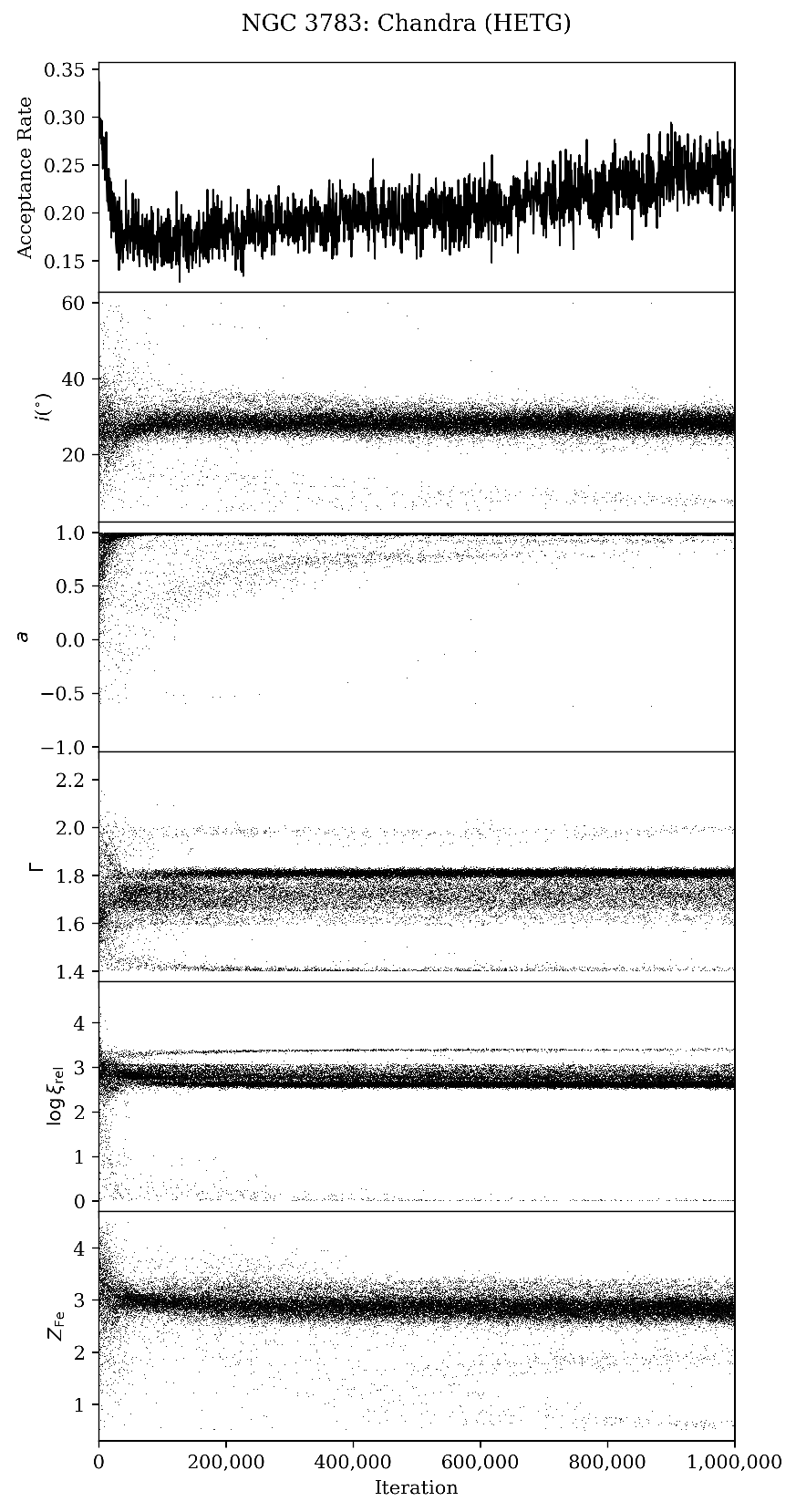}%
\end{center}
\caption{Top panel: The acceptance rates in MCMC sampling plotted against the iteration number for all datasets, group 1, and group 2. 
Bottom panels: Markov chain trace plots of the model parameters over iterations where updates occurred.
The parameters are the inclination angle ($i$[$^{\circ}$]), the spin parameter ($a$), the source photon index ($\Gamma$), the ionization parameter of \textsf{relxillCp} ($\log \xi_{\rm rel}$ [erg\,cm\,s$^{-1}$]), and the iron abundance ($Z_{\rm Fe}$ [$Z_{\odot}$]). The figure set includes the plots for three datasets (all, Group 1, and Group 2) fitted with the model.
\newline
(The complete figure set (3 images) is available.) 
\label{ngc3783:fig:chain:1}
}

\figsetstart
\figsetnum{4}
\figsettitle{The acceptance rates and Markov chain trace plots of the model parameters fitted to \textit{Chandra} HETGS datasets (all, Group 1, and Group 2) of NGC\,3783.}

\figsetgrpstart
\figsetgrpnum{4.1}
\figsetgrptitle{All data.}
\figsetplot{figures/fig4_ngc3783_chandra_heg_model4_relxillCp_mcmc_chain.eps}
\figsetgrpnote{The acceptance rates and Markov chain trace plots for all the \textit{Chandra} HETGS data of NGC\,3783.}
\figsetgrpend

\figsetgrpstart
\figsetgrpnum{4.2}
\figsetgrptitle{Group 1.}
\figsetplot{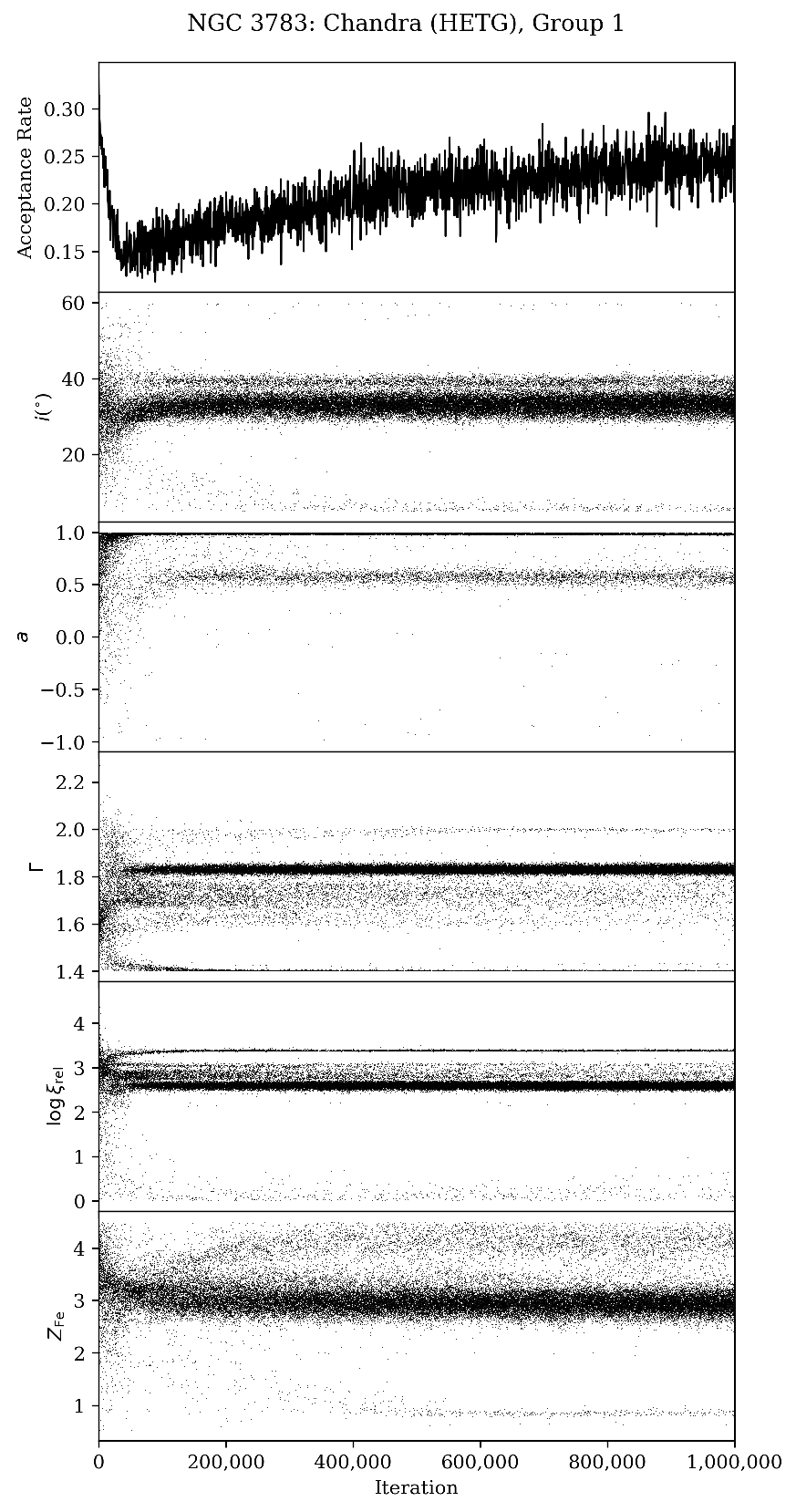}
\figsetgrpnote{The acceptance rates and Markov chain trace plots for the \textit{Chandra} HETGS Group 1 of NGC\,3783.}
\figsetgrpend

\figsetgrpstart
\figsetgrpnum{4.3}
\figsetgrptitle{Group 2.}
\figsetplot{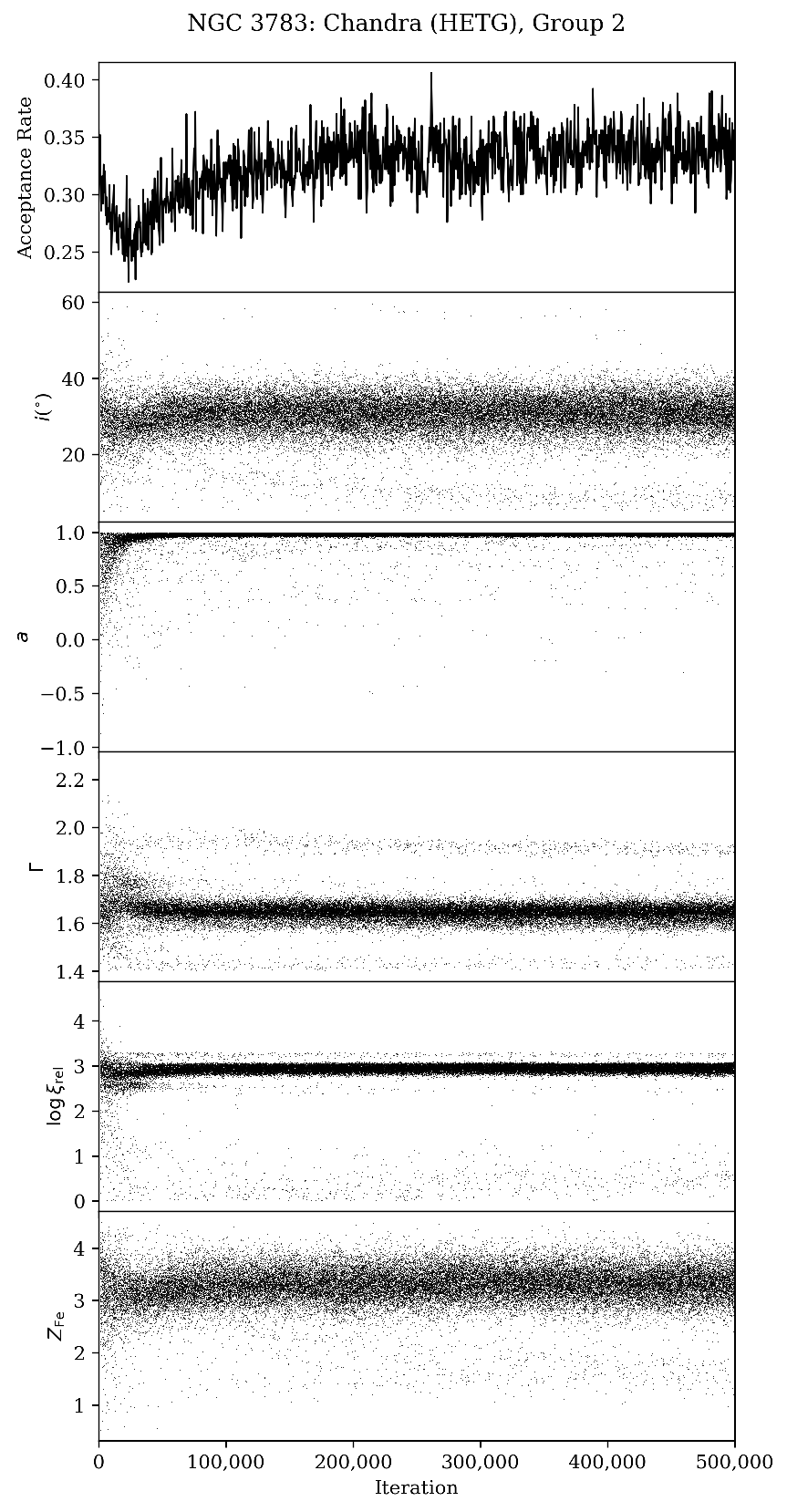}
\figsetgrpnote{The acceptance rates and Markov chain trace plots for  the \textit{Chandra} HETGS Group 2 of NGC\,3783.}
\figsetgrpend

\figsetend

\end{figure}

\begin{table*}
\begin{center}
\caption[]{Best-fit Parameters for the Spectral Model.
\label{ngc3783:model:params}}
\footnotesize
\begin{tabular}{lllll}
\hline\hline\noalign{\smallskip}
Component & Parameter & All datasets & high/soft (Group\,1)  & low/hard (Group\,2) \\
\noalign{\smallskip}
\hline\noalign{\smallskip} 
\multicolumn{5}{c}{Model: \textsf{constant}\,$\times$\,\textsf{tbnew}\,$\times$\,\textsf{wa1}\,$\times$\,(\textsf{relxillCp}\,$+$\,\textsf{xillverCp})} \\
\noalign{\smallskip}
\hline\noalign{\smallskip}
\textsf{tbnew} & $N_{\rm H}$ ($10^{20}$\,cm$^{-2}$) \dotfill & ${9.91}$ (\textit{f}) &  ${9.91}$ (\textit{f}) &  ${9.91}$ (\textit{f}) \\
\noalign{\smallskip}
\textsf{wa1} & $N_{\rm H}$ ($10^{22}$\,cm$^{-2}$) \dotfill & ${3.00}_{-0.16}^{+0.17}$ &  ${3.04}_{-0.17}^{+0.16}$ &  ${3.15}_{-0.29}^{+0.13}$ \\
             & $\log \xi$ (erg\,cm\,s$^{-1}$) \dotfill & ${2.23}_{-0.02}^{+0.02}$ &  ${2.23}_{-0.02}^{+0.02}$ &  ${1.85}_{-0.04}^{+0.04}$ \\
             & $v_{\rm out}$ (km\,s$^{-1}$) \dotfill & $ { -410 }_{ -20 }^{ +30 } $ &  $ { -410 }_{ -20 }^{ +40 } $ &  $ { -370 }_{ -80 }^{ +90 } $ \\
\noalign{\smallskip}
\textsf{relxillCp} & $K_{\rm rel}$ ($10^{-5}$) \dotfill & ${12.52}_{-0.05}^{+0.05}$ &  ${14.36}_{-0.06}^{+0.06}$ &  ${7.57}_{-0.08}^{+0.08}$ \\
                   & $i$ ($^{\circ}$) \dotfill & ${28}_{-5}^{+8}$ (m) &  ${33}_{-5}^{+9}$ (m) &  ${31}_{-10}^{+9}$ (m) \\
                   & $a$ \dotfill & ${0.98}_{-0.12}^{+0.02}$ (m) &  ${0.99}_{-0.42}^{+0.01}$ (m) &  ${0.98}_{-0.11}^{+0.02}$ (m) \\
                   & $R_{\rm in}$ ($R_{\rm ISCO}$) \dotfill & $ { 1 } $ (\textit{f}) &  $ { 1 } $ (\textit{f}) &  $ { 1 }$ (\textit{f}) \\
                   & $R_{\rm out}$ ($R_g$) \dotfill & $ { 400 } $ (\textit{f}) &  $ { 400 }$ (\textit{f}) &  $ { 400 } $ (\textit{f}) \\
                   & $R_{\rm br}$ ($R_g$) \dotfill & $(R_{\rm out}-1)/2+1$ &  $(R_{\rm out}-1)/2+1$ &  $(R_{\rm out}-1)/2+1$ \\
                   & $q_{\rm in}$ \dotfill & ${6.47}_{-0.24}^{+0.27}$ &  ${7.34}_{-0.29}^{+0.32}$ &  ${5.46}_{-0.32}^{+0.39}$ \\
                   & $\Gamma$ \dotfill & ${1.81}_{-0.41}^{+0.01}$ (m) &  ${1.83}_{-0.43}^{+0.02}$ (m) &  ${1.65}_{-0.25}^{+0.09}$ (m) \\
                   & $\log \xi_{\rm rel}$ (erg\,cm\,s$^{-1}$) \dotfill & ${2.62}_{-0.07}^{+0.78}$ (m) &  ${2.60}_{-0.11}^{+0.80}$ (m) &  ${2.97}_{-0.33}^{+0.35}$ (m) \\
                   & $\log n$ (cm$^{-3}$) \dotfill & $ { 15 } $  (\textit{f}) &  $ { 15 } $  (\textit{f}) &  $ { 15 } $  (\textit{f}) \\
                   & $Z_{\rm Fe}$ ($Z_{\odot}$) \dotfill & ${2.9}_{-0.6}^{+0.9}$ (m) &  ${3.0}_{-0.3}^{+1.5}$ (m) &  ${3.3}_{-0.8}^{+0.7}$ (m) \\
                   & k$T_{\rm e}$ (keV) \dotfill & ${60.0}_{-1.0}^{+1.0}$ &  ${60.0}_{-1.1}^{+1.1}$ &  ${60.0}_{-2.5}^{+2.5}$ \\
                   & $f_{\rm refl,rel}$ \dotfill & ${6.51}_{-0.04}^{+0.04}$ &  ${6.67}_{-0.04}^{+0.04}$ &  ${7.31}_{-0.12}^{+0.12}$ \\
\noalign{\smallskip} 
\textsf{xillverCp} & $K_{\rm xil}$ ($10^{-5}$) \dotfill & ${11.40}_{-0.70}^{+0.71}$ &  ${12.04}_{-0.85}^{+0.86}$ &  ${13.7}_{-1.5}^{+1.4}$ \\
                   & $\log \xi_{\rm xil}$ (erg\,cm\,s$^{-1}$) \dotfill & ${0.00}_{}^{+1.50}$ &  ${0.00}_{}^{+1.51}$ &  ${2.13}_{-0.09}^{+0.09}$ \\
                   & $v_{\rm shift}$ (km\,s$^{-1}$) \dotfill & $ { 620 }_{ -70 }^{ +80 } $ &  $ { 610 }_{ -80 }^{ +90 } $ &  $ { 810 }_{ -160 }^{ +130 } $ \\
                   & $f_{\rm refl,xil}$ \dotfill & $ { -1 }$ (\textit{f}) &  $ { -1 }$ (\textit{f}) &  $ { -1 }$ (\textit{f}) \\
\noalign{\smallskip} 
\textsf{constant} & $C_{\text{high/soft}}$ \dotfill & $ { 1.31 } $ & $ { 1.15 } $ & -- \\
                  & $C_{\text{low/hard}}$ \dotfill & $ { 0.76 } $ & -- & $ { 1.30 } $ \\
\noalign{\smallskip} 
Statistic & $\chi^2/{\rm d.o.f}$ \dotfill &  $ 3085 $/$ 3656 $ ($ 0.844 $) &  $ 1557 $/$ 1821 $ ($ 0.855 $) &  $ 1277 $/$ 1821 $ ($ 0.701 $) \\
          & Data bins \dotfill & $ 3670 $ &  $ 1835 $ &  $ 1835 $\\ 
\noalign{\smallskip}
          & Counts \dotfill & 222,471  &  184,112 &  38,359 \\
\noalign{\smallskip} 
\hline\noalign{\smallskip} 
\noalign{\smallskip} 
\end{tabular}
\end{center}
\begin{tablenotes}
\footnotesize
\item[1]\textbf{Notes.} The relativistic reflection models are in the rest frame ($z=0.009755$). The excess velocity $v_{\rm shift}$ of the non-relativistic Fe K$\alpha$ line (\textsf{xillverCp}) and the outflow velocity $v_{\rm out}$ of the warm absorber (\textsf{wa1}) are with respect to the rest frame. The confidence levels of the parameters are at 90\% confidence, whereas those determined from the MCMC chains denoted by ``(m)'' correspond to 90\% HDI of the modes. The fixed parameters denoted by ``(\textit{f})'' adopt typical values. The \textsf{tbnew} parameter is the Galactic column density ($N_{\rm H}$) from the UKSSDC. A negative unit value of the reflection fraction parameter ($f_{\rm refl,xil}$) means that it is associated only with reflection, without any direct continuum. The total number of counts corresponds to the energy range of 2--7.5 keV in the rest frame.
\end{tablenotes}
\end{table*}

In Figure~\ref{ngc3783:fig:chain:1}, the time series of the acceptance rates in MCMC sampling are presented as a function of iterations, showing that there are burn-in phases in the early stages, but the acceptance rates achieve the convergence domain, after 400,000 iterations (150,000 in Group 2). 
Figure~\ref{ngc3783:fig:chain:1} also shows the time series of the parameter values in the Markov chains where parameter updating occurred (the online figure set contains the corresponding plots for Groups 1 and 2). Each parameter exhibits a warm-up phase during which its value fluctuates significantly until convergence is reached. In particular, there are no robust constraints when the number of iterations is less than 100,000. When the parameters in the MCMC sampling converged, the inclination angle shifted from 23$^{\circ}$ to 28$^{\circ}$. Higher sampling iterations shrink the lower limit of the spin rate, suggesting near-maximal spin with statistically significant constraints.

The effective number of iterations in the MCMC samples may be estimated using the autocorrelation time ($\tau_{\rm int}$) as mentioned by \citet{Kass1998} and discussed further in \citet[][Ch.\,1]{Congdon2010} and \citet{Goodman2010}. To estimate the effective sample size, $N_{\rm eff}=N_n/\tau_{\rm int}$, the integrated autocorrelation time $\tau_{\rm int}=1+ 2\sum_{t=1}^{n} \rho(t)$ are calculated using the normalized autocorrelation $\rho(t)=\mathcal{\hat{C}}(t)/\mathcal{\hat{C}}(0)$ at lag $t$ \citep[see][]{Sokal1997}. Here, $N_t$ is the iteration number of Markov chains at the $t$-th lag, $n$ is the total number of samples chosen at different lags (e.g., walker steps), and $\mathcal{\hat{C}}(t)$ is the $t$-th lag autocovariance defined as:
\begin{align}
\mathcal{\hat{C}}(t) = & \frac{1}{n-1} \sum_{k=1}^{n-t} \Big( X(N_k)- \bar{X} \Big) \Big( X(N_{k+t})- \bar{X} \Big),
\end{align}
where $X(N_k)$ is the sampling value in the iteration number $N_k$ of lag $k$ for the parameter of interest, and $\bar{X}=\sum_{k=1}^{N_n} X(N_k)/N_n$ is the mean of all samples for the given parameter. To obtain the effective sample size for each parameter, we smoothed the samples using the Savitzky-Golay filter \citep{Savitzky1964} with a window size of 500 iterations and a polynomial order of 4. We then determined the effective sample sizes of the free parameters in Bayesian statistics ($i$, $a$, $\Gamma$, $\log \xi_{\rm rel}$, and $Z_{\rm Fe}$) from the samples updated at walker-step lags, which are listed in Table~\ref{ngc3783:model:eff:num}. It can be seen that the spin parameter ($a$) converges by 410,500 iterations for the model fitted to all the data, which can be easily noticed in the Markov chain trace plot in Figure~\ref{ngc3783:fig:chain:1}. We see that approximately 500,000 iterations may be sufficient for effective MCMC sampling of the free parameters in all the datasets and Group 1, whereas a lower number of around 250,000 iterations could be required for   Group 2. Furthermore, the trend in the acceptance ratio time series toward the convergence region (0.2--0.5) suggests the same number of iterations.

\begin{table}
\begin{center}
\caption[]{Effective Sample Sizes in Bayesian Statistics.
\label{ngc3783:model:eff:num}}
\footnotesize
\begin{tabular}{lrrr}
\hline\hline\noalign{\smallskip}
Parameter & \multicolumn{1}{c}{All} & \multicolumn{1}{c}{Group\,1}  & \multicolumn{1}{c}{Group\,2} \\
\noalign{\smallskip}
\hline\noalign{\smallskip} 
$i$($^{\circ}$) & 546,900  & 445,900 & 165,000  \\
\noalign{\smallskip} 
$a$ & 410,500 & 317,500 & 257,500 \\
\noalign{\smallskip} 
$\Gamma$ & 669,100 & 241,400  & 213,600 \\
\noalign{\smallskip} 
$\log \xi_{\rm rel}$ & 476,900 & 442,100  & 109,200 \\
\noalign{\smallskip} 
$Z_{\rm Fe}$ & 365,100 & 489,500 & 210,200 \\
\noalign{\smallskip} 
\hline\noalign{\smallskip} 
\noalign{\smallskip} 
\end{tabular}
\end{center}
\end{table}

\setcounter{figure}{4}
\begin{figure*}
\begin{center}
\includegraphics[width=0.92\textwidth, trim = 0 0 0 0, angle=0]{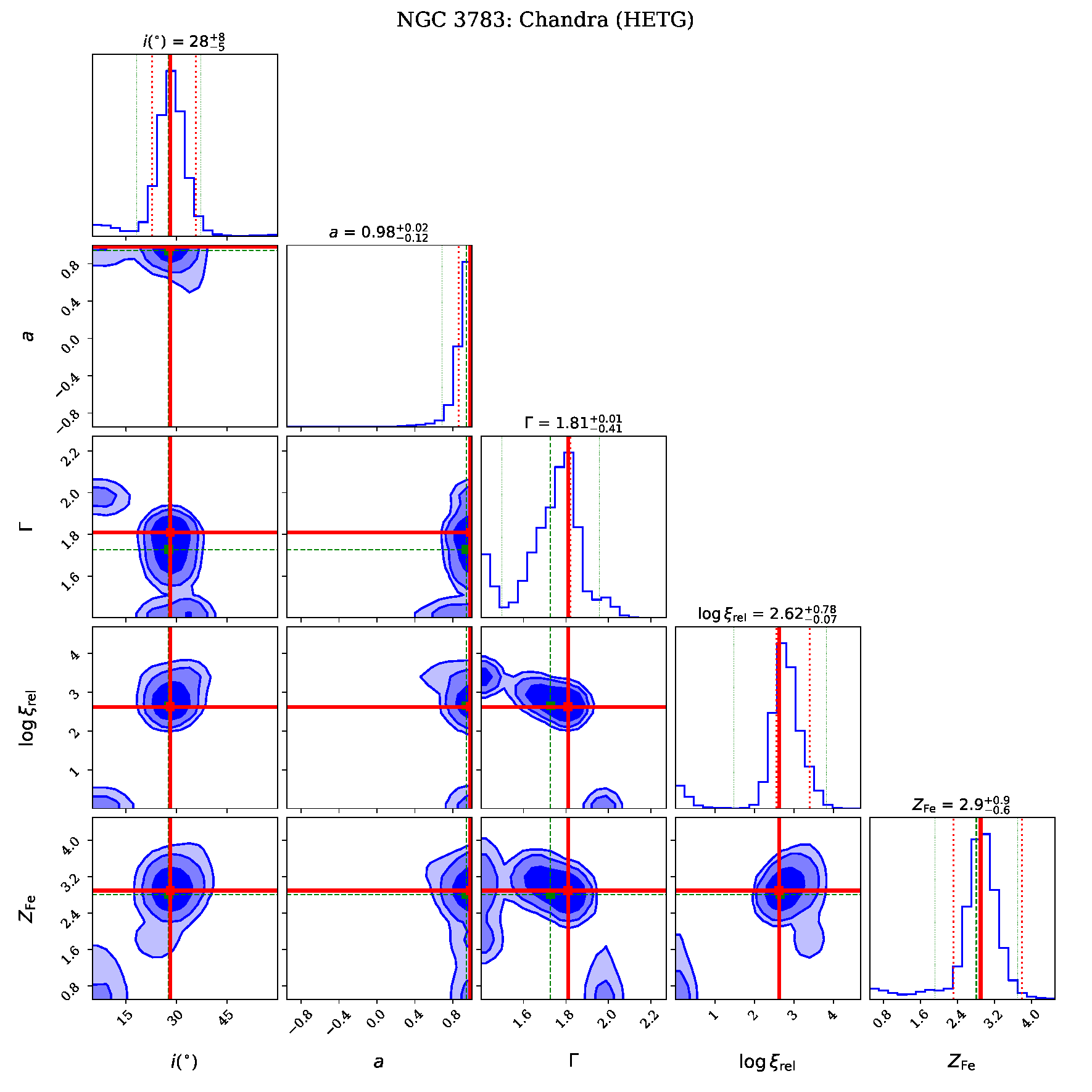}%
\end{center}
\caption{The posterior probability distributions of the model parameters fitted to all the \textit{Chandra} ACIS-S/HETGS data ($m=\pm1$ orders; HEG) of NGC\,3783. Confidence contours are plotted at the 1-$\sigma$ (68\%), 2-$\sigma$ (95\%), and 3-$\sigma$ (99.7\%) levels. The locations of the modes and means are shown by the red color solid and green color dashed lines, respectively. The 90\% HDI levels of the modes and the 1.645-$\sigma$ (90\%) standard deviation limits of the means are also plotted by the red color and green color dotted lines in the probability density function plots, respectively. The parameters are as follows: the inclination angle ($i$[$^{\circ}$]), the spin parameter ($a$), the source photon index ($\Gamma$), the ionization parameter of \textsf{relxillCp} ($\log \xi_{\rm rel}$ [erg\,cm\,s$^{-1}$]), and the iron abundance ($Z_{\rm Fe}$ [$Z_{\odot}$]). The figure set includes the plots for three datasets (all, Group 1, and Group 2) fitted with the spectral model.
\newline
(The complete figure set (3 images) is available.) 
\label{ngc3783:fig:mcmc:1}
}

\figsetstart
\figsetnum{5}
\figsettitle{The posterior constraints on the model parameters fitted to \textit{Chandra} HETGS datasets (all, Group 1, and Group 2) of NGC\,3783. Confidence contours are plotted at the 1-$\sigma$, 2-$\sigma$, and 3-$\sigma$ levels.}

\figsetgrpstart
\figsetgrpnum{5.1}
\figsetgrptitle{All data.}
\figsetplot{figures/fig5_ngc3783_chandra_heg_model4_relxillCp_mcmc.eps}
\figsetgrpnote{The posterior constraints for all the \textit{Chandra} HETGS data of NGC\,3783.}
\figsetgrpend

\figsetgrpstart
\figsetgrpnum{5.2}
\figsetgrptitle{Group 1.}
\figsetplot{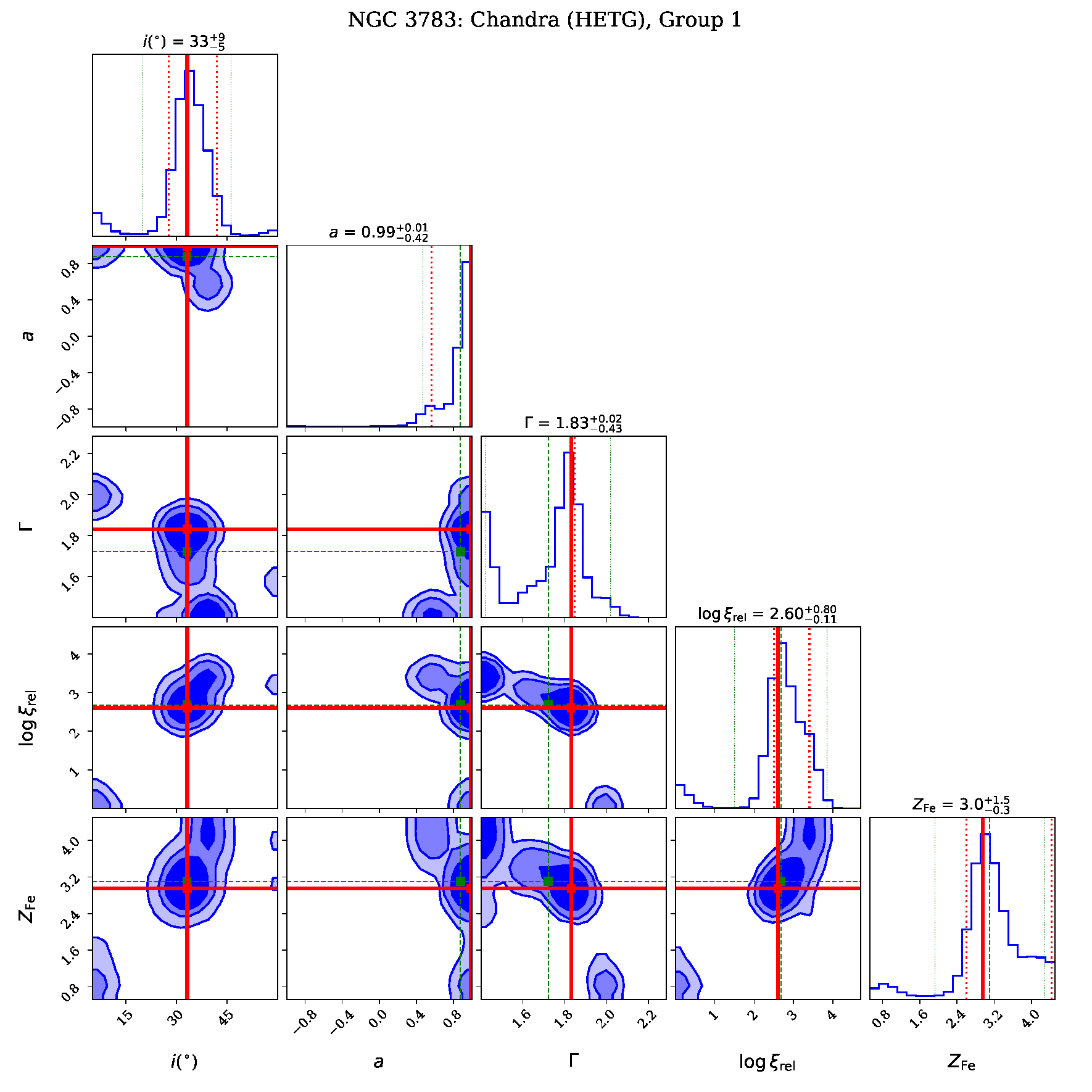}
\figsetgrpnote{The posterior constraints for the \textit{Chandra} HETGS Group 1 of NGC\,3783.}
\figsetgrpend

\figsetgrpstart
\figsetgrpnum{5.3}
\figsetgrptitle{Group 2.}
\figsetplot{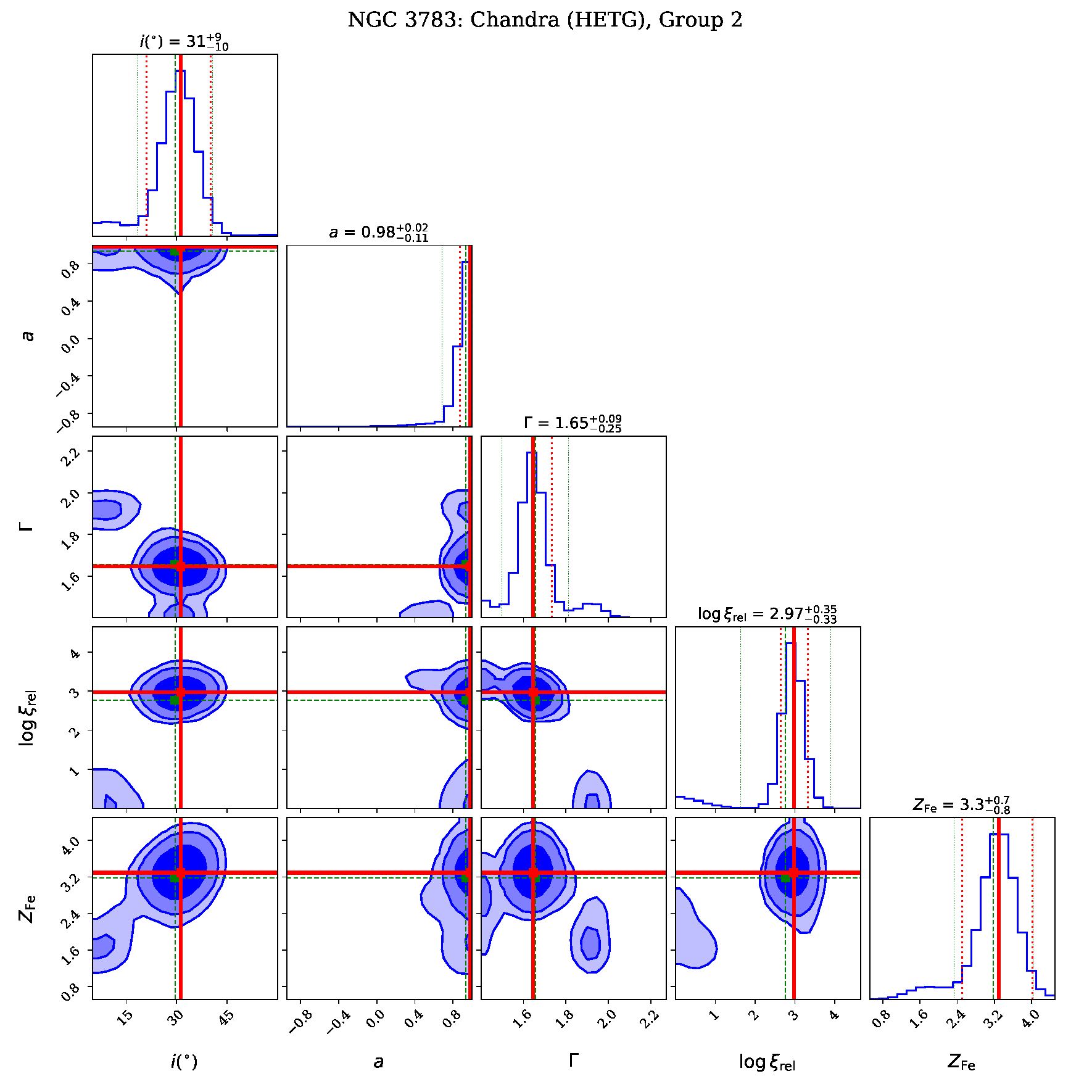}
\figsetgrpnote{The posterior constraints for the \textit{Chandra} HETGS Group 2 of NGC\,3783.}
\figsetgrpend

\figsetend

\end{figure*}

We used the Python package \textsf{corner} \citep{Foreman-Mackey2016} to map the posterior probability distributions of the fitting parameters. 
Figure~\ref{ngc3783:fig:mcmc:1} presents the posterior constraints on the model parameters fitted to all the first-order HEG data. It shows the 1-$\sigma$, 2-$\sigma$, and 3-$\sigma$ confidence contours of the parameters versus each other (the corresponding posterior constraints for Groups 1 and 2 can be found in the online figure set). The modes and means of the best-fitting parameters are also plotted by solid (red) and dashed (green) lines, respectively. Moreover, the probability density function (PDF) plots also show the 90\% highest density intervals (HDI) of the modes (red dotted lines) and the 1-$\sigma$ standard deviation levels of the means (green dotted lines). The modes and HDI levels were obtained using the kernel-density estimation (KDE) function for the Gaussian distribution from the \textsf{SciPy} statistical library \citep{Virtanen2020} and the HDI task of \textsf{ArviZ} \citep{Kumar2019}, respectively. Table~\ref{ngc3783:model:params} presents the mode values and confidence ranges with 90\% HDI for the Bayesian constraints on the parameters, along with the values of the best-fitting parameters determined from our spectral modeling (\S\,\ref{ngc3783:spec}).

As seen in Figure~\ref{ngc3783:fig:mcmc:1} for all the HETGS data, there are robust posterior constraints on the spin parameter ($a$), inclination ($i$), photon index ($\Gamma$), iron abundance ($Z_{\rm Fe}$), and ionization parameter of \textsf{relxillCp} ($\log \xi_{\rm rel}$). The lower limit of the photon index is more uncertain than the upper limit in the models for all data and Group 1 (high/soft). Although the likelihood peak for the inclination angle in Group 2 is approximately similar to those obtained for all the data and Group 1, it is more uncertain owing to the lower available counts. The PDF plots of the spin parameter ($a$) indicate a near-maximal spin of $\geqslant 0.86$ at 90\% confidence for all the data, which is identical to those derived in other dataset groups, albeit with a more uncertain lower limit in Group 1. We caution that Group 2 with a low number of counts could ($3.8 \times 10^{4}$; see Table~\ref{ngc3783:obs:log}) result in weak constraints on some parameters of the relativistic reflection model. To put robust constraints on relativistic emission, we may need to have $> 1.5 \times 10^{5}$ counts in the energy range of 2--10\,keV. We see that the ionization parameter ($\log \xi_{\rm rel}$) of the innermost region (\textsf{relxillCp}) is higher than that of the distant region \textsf{xillverCp}, which is explained by the fact that the inner regions are closer to the luminous ionizing radiation of the corona.

Our constrained spin parameter and inclination are consistent within their corresponding confidence ranges for all the datasets. The modes of the posterior distribution for the inclination are around $30^{\circ}$ in different dataset groups. Our spin values are in agreement with those reported by \citet{Brenneman2011} and \citet{Reynolds2012}. Our derived inclination of $28^{\circ}{}^{+8}_{-5}$ is slightly higher than the previous values: $19$--$22^{\circ}$ \citet{Brenneman2011} and $22$--$24^{\circ}$ \citep{Reynolds2012}.
Recently, the introduction of GRAVITY \citep{GRAVITYCollaboration2017}, a next-generation instrument at the Very Large Telescope Interferometer (VLTI), has enabled significant advancements in AGN research, which has allowed the detection of broad-line regions (BLRs) in several AGNs \citep[e.g.,][]{GravityCollaboration2018}. A rotating BLR model fitted to the VLTI/GRAVITY observations of NGC\,3783 implies an inclination of either $23^{\circ}{}^{+16}_{-10}$ \citep{GRAVITYCollaboration2021a} or $32^{\circ}\pm4$ \citep{GRAVITYCollaboration2021b}, which is in agreement with the confidence limits of our values obtained for the inclination angle of the accretion disk.



\section{Discussion}
\label{ngc3783:discussion}

We leveraged all available \textit{Chandra} ACIS-S/HETGS data for the Seyfert 1 galaxy NGC\,3783 to investigate the X-ray relativistic reflection from the accretion disk. 
Deep \textit{Suzaku} studies indicated a near-maximal spin parameter of at least $0.89$ \citep{Brenneman2011,Reynolds2012}, while other analyses of \textit{Suzaku} data pointed to a slow spin of $0.24$ \citep{Patrick2012} and even a slightly negative spin of $-0.04$ \citep{Patrick2011}. 
It has been speculated that iron abundance may be responsible for the different values derived for the black-hole spin in this object. \citet{Reynolds2012} argued that the iron abundances in the inner accretion disk have important implications for spin measurement, and demonstrated the necessity of an iron abundance of $2$--$4Z_{\odot}$ and a spin of $a>0.89$. The previous spectral analysis conducted by \citet{Brenneman2011} also obtained similar results, so the discrepancy in the spin was attributed to the assumption of a fixed solar abundance ($Z_{\odot}$). Assuming a disk that is chemically uniform and has the same amount of iron in the farthest and closest parts, our model also supports an iron-rich chemistry with $Z_{\rm Fe}/Z_{\odot}= { 2.9}_{ -0.6 }^{ +0.9 }$ for all the HEG data put together. This super-solar metallicity aligns well with what has been deduced from optical/UV observations of the BLRs of type 1 AGNs \citep[e.g.,][]{Warner2004,Fan2023}, as well as the Fe abundance found in their X-ray spectral analyses \citep[e.g.,][]{deRosa2007,Beuchert2017}. In particular, our Bayesian posterior constraints demonstrate that we can constrain iron abundance with statistically significant confidence. 

\subsection{Relativistic Reflection Emission}
\label{ngc3783:relativistic}

This study focuses on the measurement of relativistic reflection in the Seyfert 1 AGN NGC\,3783. In our spectral models implemented with \textsf{relxillCp}, the irradiation resulting in the illuminated inner disk was simply described by an emissivity law with a powerlaw profile ($r^{-q}$). In addition, the ionizing source, which directly illuminates the inner disk, has a spectral photon index of $\approx 1.8^{+0.0}_{-0.4}$ in all the data and during the high/soft state, whereas there is a lightly harder photon index ($\Gamma \approx 1.7^{+0.1}_{-0.3}$) in the low/hard state. The ionization parameter that characterizes the inner relativistic zone of the accretion disk shows a more ionized ($\log \xi_{\rm rel} \sim 2.6$--$3$) than that of the distant layer of the disk ($\log \xi_{\rm xil} \sim 0$--2). We also see that the ionization parameter in \textsf{relxillCp} is slightly higher in the low/hard state. 

Our models provide physically plausible power-law spectral indices, as well as robust constraints on spin and inclination, namely $a=0.98^{+0.02}_{-0.12}$ and $i=28^{\circ}{}^{+8}_{-5}$ (all datasets). In particular, the lower limit of the spin is in agreement with $a \geqslant 0.88$ (99\% confidence) obtained by \citet{Brenneman2011} using \textit{Suzaku} XIS+PIN data, in addition to $a > 0.89$ from MCMC analysis of \textit{Suzaku} data by \citet{Reynolds2012}. As the \textit{Chandra} telescope collects only photons in the energy range below 10\,keV, the reflection fraction constrained by our models could be more uncertain than those obtained with \textit{Suzaku} XIS+PIN data covering energies up to 40\,keV. An inadequate constraint on the reflection fraction could potentially weaken the Bayesian constraints on the inclination, particularly in Group 2 with low counts. Even without Compton reflection humps ($>10$\,keV), our study demonstrates that the spin can still be determined from the multi-epoch \textit{Chandra} HETGS data with substantial photons.

\subsection{Narrow Iron K$\alpha$ Line}
\label{ngc3783:narrowline}

The detection of the narrow Fe K$\alpha$ line component using the \textit{Chandra} ACIS-S/HETGS with high spectral resolution helps us to better interpret the distant regions that reflect the coronal radiation in the Seyfert 1 galaxy NGC\,3783. The \textit{Suzaku} XIS detectors seemed incapable of resolving the spectral features that were captured using \textit{Chandra} grating spectroscopy. The time-averaged \textit{Chandra} spectrum shows a narrow, bright Fe K$\alpha$ line, which mostly originates from regions close to the center. There is a blue wing beyond the narrow line, which, according to our first model, is part of the relativistic innermost region (see Fig.\,\ref{ngc3783:fig:model}). 

The peak of the iron K$\alpha$ line was found to be located at an energy lower than that expected from the rest frame, resulting in an excess velocity of $620^{+80}_{-70}$\,km\,s$^{-1}$ with respect to the host galaxy. Similarly, an excess velocity of 326\,km\,s$^{-1}$ was measured using the \textit{Chandra} ACIS-S/HETGS observations of Centaurus\,A, which was ascribed to a warped compact disk on subparsec scales \citep{Bogensberger2024}. It was also suggested that the asymmetric shape of the narrow Fe K$\alpha$ line in NGC\,4151 seen with \textit{Chandra} HETGS was caused by a warp or transition region within the disk that exposes more neutral gas in the central part \citep{Miller2018}. Recent \textit{XRISM} spectroscopic observations of NGC\,4151 also suggest contributions from the innermost BLR (optical/X-ray) to the Fe K$\alpha$ line, in addition to a potentially warped accretion disk \citep{XRISMCollaboration2024}. 
Specifically, \citet{GRAVITYCollaboration2021a} determined that the BLR in NGC\,3783 can be accurately characterized as a thick, rotating disk with the highest cloud concentration in the inner region. Alternatively, as proposed by \citet{Miller2018}, a failed wind could also result in an asymmetric narrow Fe K$\alpha$ line. In particular, our analysis was conducted using multiple \textit{Chandra} data; therefore, a temporary transition region or a clumpy failed wind that occurred in some epochs could also contribute to the excess redshift of the narrow Fe K$\alpha$ emission.

\subsection{Reflection Features in Different Spectral States}
\label{ngc3783:state}

The hardness-ratio diagrams of NGC 3783 in Fig.~\ref{ngc3783:fig:hdr:1} show transitions between two different spectral states, ``low/hard'' and ``high/soft'', although they are dissimilar to those with the same names observed in black hole binary (BHB) systems \citep[see review by][]{Remillard2006}. While certain patterns are evident in the hardness-ratio diagrams, they lack sufficient strength to be linked with any spectral-state transitions comparable to those observed in BHBs \citep[see, e.g., Fig.\,1 in][]{Homan2005}. In the case of NGC 3783, transient obscuration events are thought to occur, which are explained by the eclipse of the X-ray ionizing continuum caused by line-of-sight dense material \citep{Mehdipour2017,Kaastra2018,Kriss2019,DeMarco2020}. However, spectral-state transitions in BHBs are believed to be related to changes in the mass accretion rate \citep[e.g.,][]{Esin1997}, which are dissimilar to those caused by obscuration events in AGNs.

As seen in Table~\ref{ngc3783:model:params}, the spectral index ($\Gamma$) is slightly higher (softer) in the high/soft state, although it is within the confidence limits of that in the low/hard state. Similarly, the ionization parameter of the inner regions ($\log \xi_{\rm rel}$) is slightly lower in the soft state, but consistent with the confidence intervals in the hard state. However, the ionization parameter of distant regions ($\log \xi_{\rm xil}$) appears to be lower in the high/soft state, albeit with a highly uncertain upper limit. In addition, the normalization factor of the \textsf{relxillCp}  component in the bright/soft state is nearly twice that in the faint/hard state, which may be explained by the fact that continuum emission is more obscured in the hard state owing to the transient obscuration events.
Considering the values obtained for the cross-normalization constant, multiplying $C_{\text{high/soft}}$ and $C_{\text{low/hard}}$ of the model fitted to all the datasets by the normalization of \textsf{relxillCp} yield approximately the same values derived from the multiples of the \textsf{relxillCp} normalization factors and constant values separately derived for Groups 1 and 2. The normalization factor of the \textsf{xillverCp} component is slightly higher in the low state. In particular, NGC 4151 exhibited an asymmetric narrow Fe K$\alpha$ line \citep{Miller2018}, which was found to contain contributions from the innermost BLR \citep{XRISMCollaboration2024}. GRAVITY observations of NGC 3783 also revealed the BLR, which was well modeled by a thick disk with a radial cloud distribution peaking in the inner region \citep{GRAVITYCollaboration2021a}. Hence, the higher normalization of \textsf{xillverCp} in the low/hard state may be related to a higher contribution from the BLR to the  narrow Fe K$\alpha$ line when there are more obscuring materials.

We notice a marginally higher emissivity index ($q$) in the soft state, which indicates a steeper decline in the radial emissivity profile ($r^{-q}$) of the inner accretion disk. This may imply that the X-ray illumination of the inner layers of the accretion disk was slightly elevated in the bright/soft state. We also see that the reflection fraction of the inner regions ($f_{\rm refl,rel}$), which is the ratio of the coronal intensity illuminating the disk to the  observed coronal intensity \citep{Dauser2014,Dauser2016}, is slightly higher in the hard state, which may be associated with stronger relativistic reflection features in that state. The confidence ranges of iron abundance in different spectral states are generally consistent with each other; therefore, the metallicity of the accretion disk remains the same over the course of different spectral states.

\section{Conclusion}
\label{ngc3783:conclusion}
 
X-ray \textit{Chandra} observations of NGC\,3783 showed stochastic hourly small-scale variability in addition to long-term brightness/hardness transitions on yearly scales, which were ascribed to transient obscuration events \citep{Mehdipour2017,Kriss2019}. According to the \textit{Chandra} light curves, the primary X-ray source was fainter and harder (2013--2016) over approximately 17\% of the collected counts. Despite spectral transitions and stochastic variability, the substantial number of $2.2 \times 10^{5}$\,counts over 2--10\,keV with the HEG assembly on \textit{Chandra} ACIS-S/HETGS across different epochs allowed us to disentangle the relativistically smeared reflection emission of the accretion disk from the X-ray continuum of the corona close to the SMBH. However, the \textit{Chandra} energy coverage prevented us from constraining the Compton hump of relativistic reflection, potentially leading to unreliable statistical confidence in some parameters.

The time-averaged \textit{Chandra} HETGS data of NGC\,3783 validate the existence of a relativistically broadened red-wing beside the Fe K$\alpha$ line at approximately 6.4 keV, yielding a spin of $a=0.98^{+0.02}_{-0.12}$ in agreement with those previously reported with \textit{Suzaku} \citep{Brenneman2011,Reynolds2012}. Additionally, all HETGS data highlighted the narrow Fe K$\alpha$ line with an excess velocity of $620^{+80}_{-70}$\,km\,s$^{-1}$, which might originate from a warped structure on subparsec scales in the accretion disk \citep[see, e.g.,][]{Miller2018,Bogensberger2024,XRISMCollaboration2024}. In summary, this study demonstrates that the black-hole spin can be measured using large counts gathered over multi-epoch HETGS observations and is not affected by intrahour variations or long-term transitions in the X-ray brightness caused by line-of-sight outflowing materials.



\begin{acknowledgments}
This research is funded by the NASA ADAP grant 80NSSC22K0626, and has made use of data obtained from the \textit{Chandra} Data Archive, 
software provided by the Chandra X-ray Center (CXC) in the application packages \textsc{ciao} and \textsf{Sherpa}, and a collection of \textsc{isis} functions (ISISscripts) provided by ECAP/Remeis observatory and MIT.
We thank the anonymous referee for constructive comments that improved our manuscript.

\end{acknowledgments}

%

\vspace{1mm}


\software{\textsc{ciao} \citep{Fruscione2006}, \textsc{xspec} \citep{Arnaud1996}, \textsc{isis} \citep{Houck2000}, \textsf{matplotlib} \citep{Hunter2007}, \textsf{numpy} \citep{vanderWalt2011}, \textsf{SciPy} \citep{Virtanen2020}, \textsf{ArviZ} \citep{Kumar2019}.}

\facilities{CXO (ACIS, HETG).}


{ \small 
\begin{center}
\textbf{ORCID iDs}
\end{center}
\vspace{-5pt}

\noindent A.~Danehkar \orcidauthor{0000-0003-4552-5997} \url{https://orcid.org/0000-0003-4552-5997}

\noindent W.~N.~Brandt \orcidauthor{0000-0002-0167-2453} \url{https://orcid.org/0000-0002-0167-2453}

}

\begin{appendix}

\section*{Supplementary Data}
\label{appendix:a}

The following figure sets are available for the electronic edition of this article:
\\
\textbf{Fig.~Set~\ref{ngc3783:fig:chain:1}.} The acceptance rates and Markov chain trace plots of the model parameters fitted to \textit{Chandra} HETGS datasets (all, Group 1, and Group 2) of NGC\,3783.
\\
\textbf{Fig.~Set~\ref{ngc3783:fig:mcmc:1}.} The posterior constraints on the model parameters fitted to \textit{Chandra} HETGS datasets (all, Group 1, and Group 2) of NGC\,3783. Confidence contours are plotted at the 1-$\sigma$, 2-$\sigma$, and 3-$\sigma$ levels.

\end{appendix}





\def\showsupplementdata{0}

\ifx\showsupplementdata\undefined

\else

\newpage


\section*{Supplementary Data} 




\FloatBarrier
\normalsize

\renewcommand{\figurename}{Fig. Set}

\figurenum{4}
\begin{figure*}
\begin{center}
\includegraphics[width=0.48\textwidth, trim = 0 0 0 0, angle=0]{figures/fig4_ngc3783_chandra_heg_model4_relxillCp_mcmc_chain.eps}%
\includegraphics[width=0.48\textwidth, trim = 0 0 0 0, angle=0]{figures/fig4_ngc3783_chandra_heg_grp1_model4_relxillCp_mcmc_chain.eps}%
\caption{The acceptance rates and Markov chain trace plots of the model parameters fitted to \textit{Chandra} HETGS datasets (all, Group 1, and Group 2) of NGC\,3783.  %
\newline
\textbf{Figure 4.1.} The acceptance rates and Markov chain trace plots for all the \textit{Chandra} HETGS data of NGC\,3783.
\newline
\textbf{Figure 4.2.} The acceptance rates and Markov chain trace plots for the \textit{Chandra} HETGS Group 1 of NGC\,3783.
\label{ngc3783:fig:chain:2}
}
\end{center}
\end{figure*}

\figurenum{4}
\begin{figure*}
\begin{center}
\includegraphics[width=0.48\textwidth, trim = 0 0 0 0, angle=0]{figures/fig4_ngc3783_chandra_heg_grp2_model4_relxillCp_mcmc_chain.eps}%
\caption{\textit{-- continued} 
\newline\hspace{\textwidth}
\textbf{Figure 4.3.} The acceptance rates and Markov chain trace plots for  the \textit{Chandra} HETGS Group 2 of NGC\,3783.
}
\end{center}
\end{figure*}

\newpage
\FloatBarrier

\figurenum{5}
\begin{figure*}
\begin{center}
\includegraphics[width=0.92\linewidth]{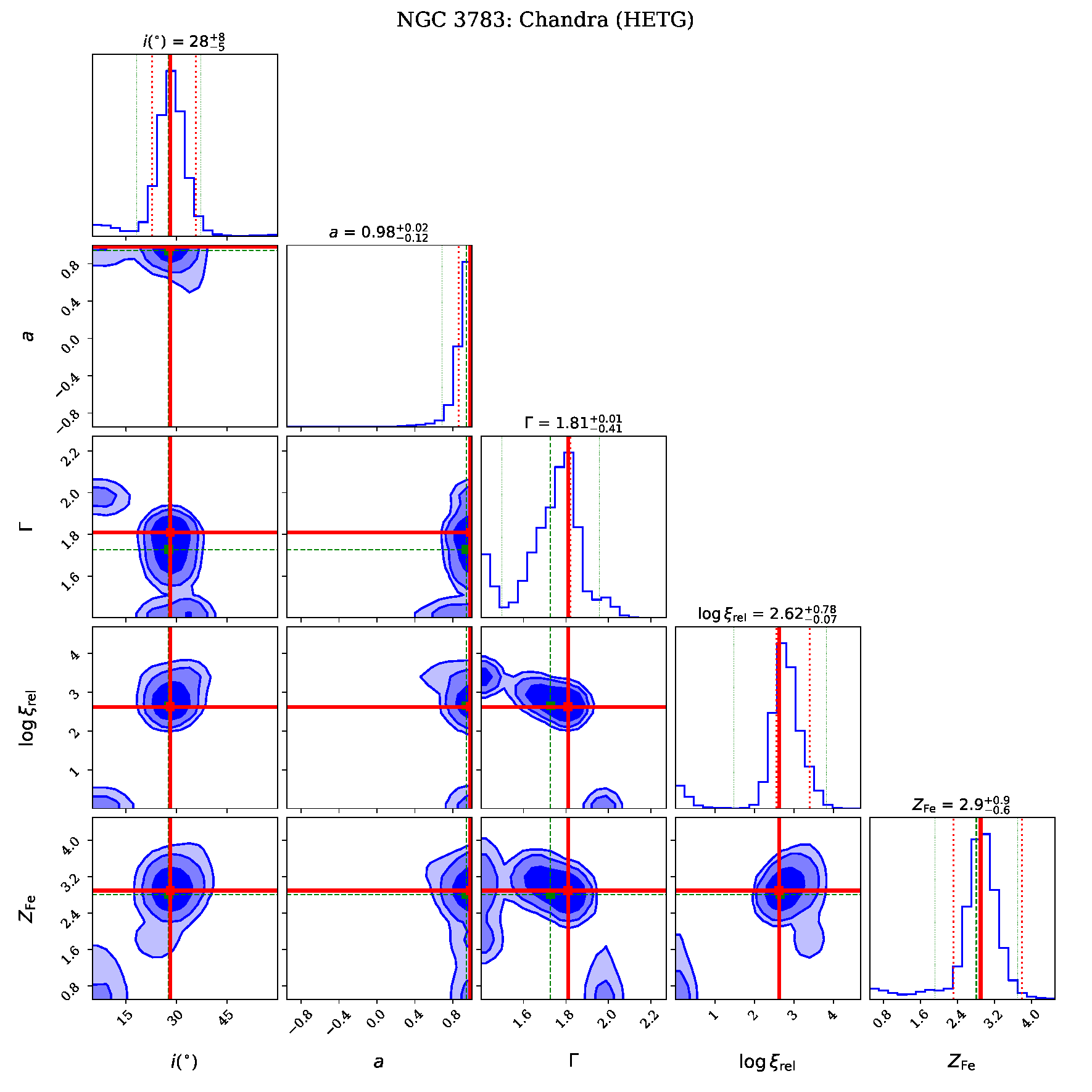}
\caption{The posterior constraints on the model parameters fitted to \textit{Chandra} HETGS datasets (all, Group 1, and Group 2) of NGC\,3783. Confidence contours are plotted at the 1-$\sigma$, 2-$\sigma$, and 3-$\sigma$ levels.  %
\newline
\textbf{Figure 5.1.} The posterior constraints for all the \textit{Chandra} HETGS data of NGC\,3783.
\label{ngc3783:fig:mcmc:2}
}
\end{center}
\end{figure*}


\figurenum{5}
\begin{figure*}
\begin{center}
\includegraphics[width=0.92\linewidth]{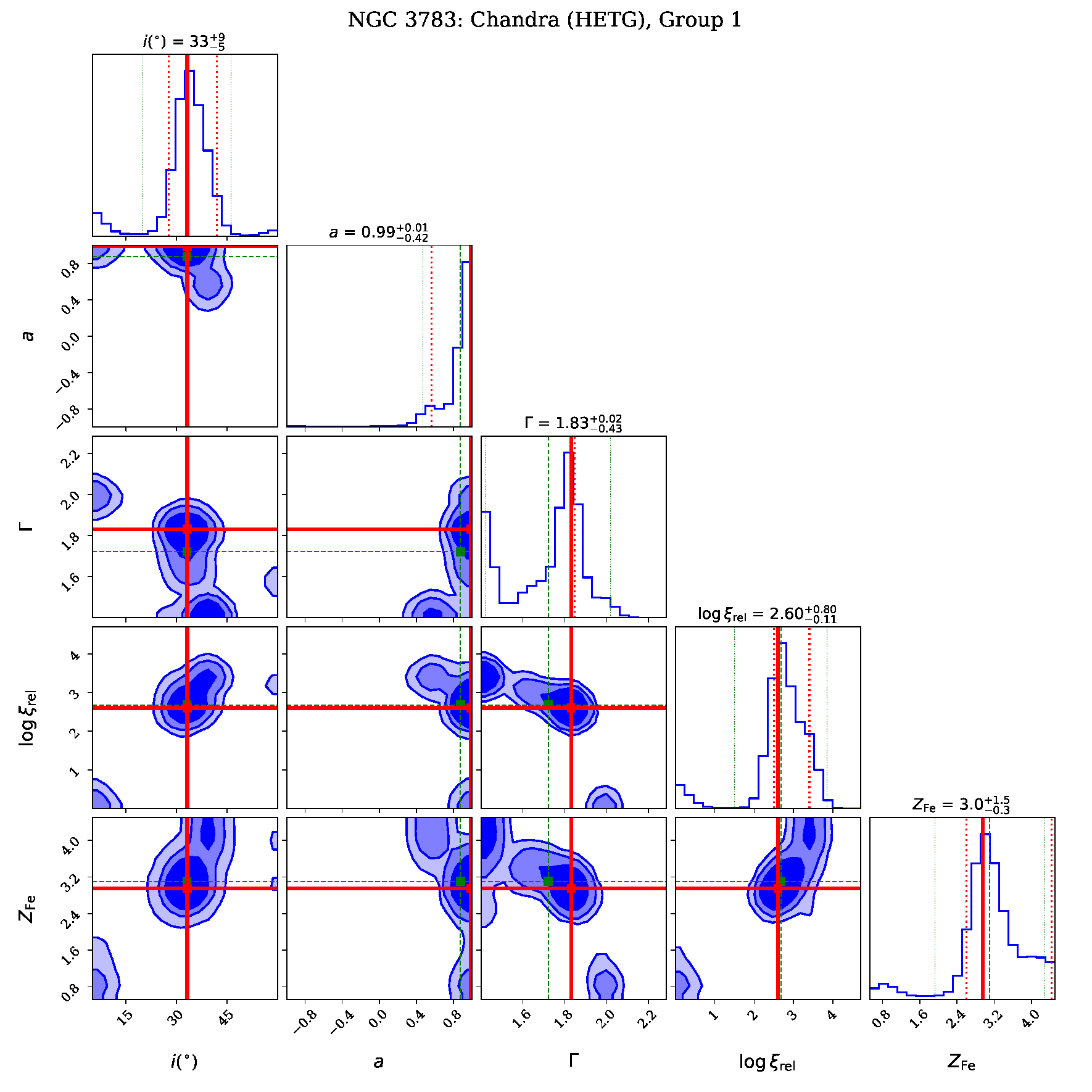}
\caption{\textit{-- continued} 
\newline\hspace{\textwidth}
\textbf{Figure 5.2.} The posterior constraints for the \textit{Chandra} HETGS Group 1 of NGC\,3783.
}
\end{center}
\end{figure*}

\figurenum{5}
\begin{figure*}
\begin{center}
\includegraphics[width=0.92\linewidth]{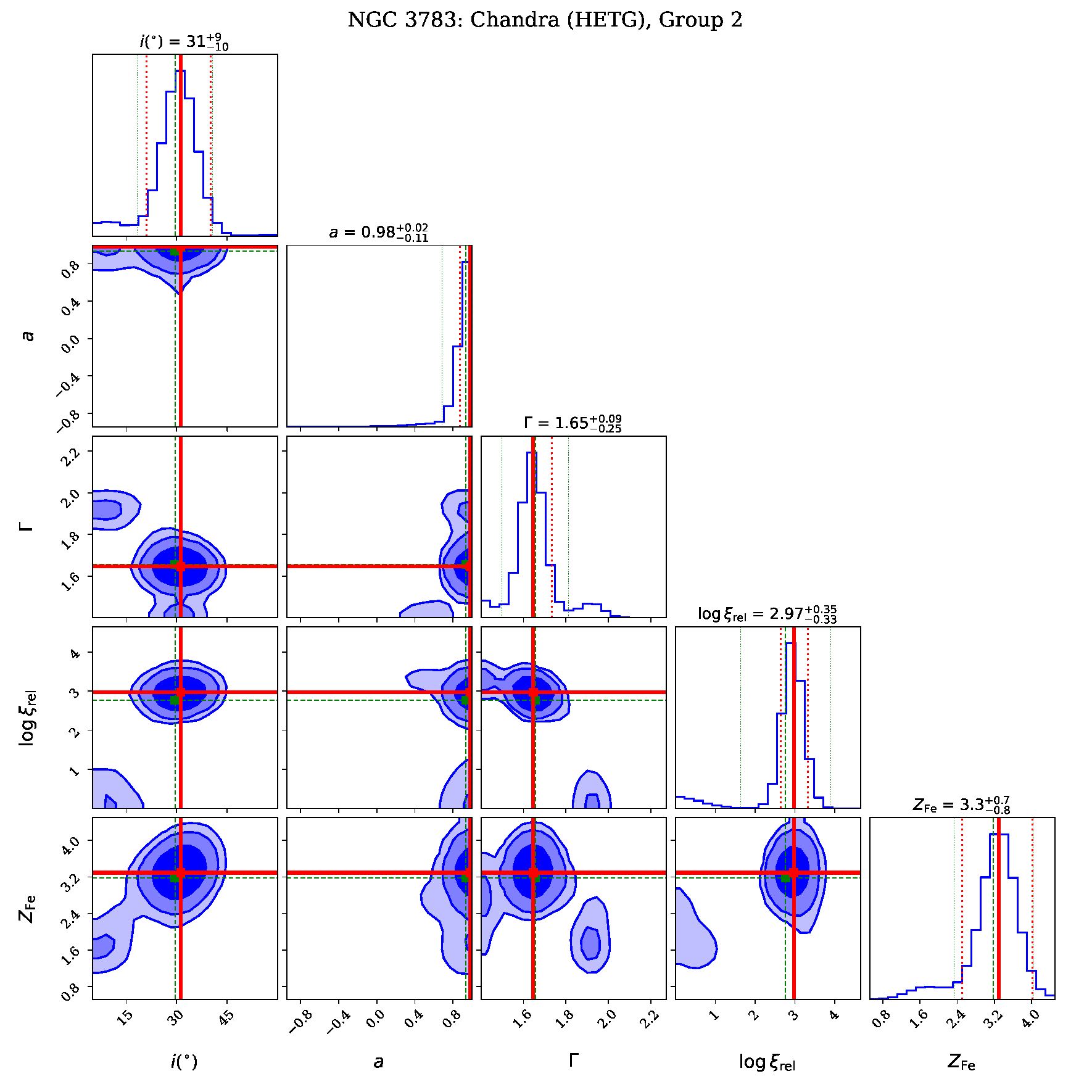}
\caption{\textit{-- continued}
\newline\hspace{\textwidth}
\textbf{Figure 5.3.} The posterior constraints for the \textit{Chandra} HETGS Group 2 of NGC\,3783.
}
\end{center}
\end{figure*}

\FloatBarrier

\fi

\end{document}